\documentclass{emulateapj}
\def\msun{$M_{\odot}$} 
\usepackage{lineno}
\usepackage{multirow}
\usepackage{subfigure}

\def\xte{{\it RXTE}} 
\def\gx{\object{GX 17+2}}

\shortauthors{Lin et al.}

\begin{document}

\title{The Spectral Evolution along the Z track of the Bright Neutron Star X-ray Binary GX 17+2}

\author{Dacheng Lin\altaffilmark{1,2}, Ronald A. Remillard\altaffilmark{3}, Jeroen Homan\altaffilmark{3}, and Didier Barret\altaffilmark{1,2}}
\altaffiltext{1}{CNRS, IRAP, 9 avenue du Colonel Roche, BP 44346, F-31028 Toulouse Cedex 4, France, email: Dacheng.Lin@irap.omp.eu}
\altaffiltext{2}{Universit\'{e} de Toulouse, UPS-OMP, IRAP, Toulouse, France}
\altaffiltext{3}{MIT Kavli Institute for Astrophysics and Space Research, MIT, 70 Vassar Street, Cambridge, MA 02139-4307, USA}

\begin{abstract}
Z sources are bright neutron-star X-ray binaries, accreting at
  around the Eddington limit. We analyze the 68 \textit{RXTE}
  observations ($\sim$270~ks) of Sco-like Z source \object{GX~17+2}
  made between 1999 October 3--12, covering a complete Z track. We
  create and fit color-resolved spectra with a model consisting of a
  thermal multicolor disk, a single-temperature-blackbody boundary
  layer and a weak Comptonized component. We find that, similar to
  what was observed for \object{XTE~J1701-462} in its Sco-like Z
  phase, the branches of \object{GX~17+2} can be explained by three
  processes operating at a constant accretion rate $\dot{M}$ into the
  disk: increase of Comptonization up the horizontal branch,
  transition from a standard thin disk to a slim disk up the normal
  branch, and temporary fast decrease of the inner disk radius up the
  flaring branch. We also model the Comptonization in an empirically
  self-consistent way, with its seed photons tied to the thermal disk
  component and corrected for to recover the pre-Comptonized thermal
  disk emission. This allows us to show a constant $\dot{M}$ along the
  entire Z track based on the thermal disk component. We also measure
  the upper kHz QPO frequency and find it to depend on the apparent
  inner disk radius $R_\mathrm{in}$ (prior to Compton scattering)
  approximately as frequency~$\propto$~$R_\mathrm{in}^{-3/2}$,
  supporting the idenfitication of it as the Keplerian frequency at
  $R_\mathrm{in}$. The horizontal branch oscillation is probably
  related to the dynamics in the inner disk as well, as both its
  frequency and $R_\mathrm{in}$ vary significantly on the horizontal
  branch but become relatively constant on the normal branch. 

\end{abstract}
\keywords{accretion, accretion disks --- starts: individual (\gx) --- stars: neutron --- X-rays: binaries --- X-rays: bursts --- X-ray: stars}

\section{INTRODUCTION}
\label{sec:intro}

\begin{figure*}
\centering
\includegraphics[width=0.99\textwidth]{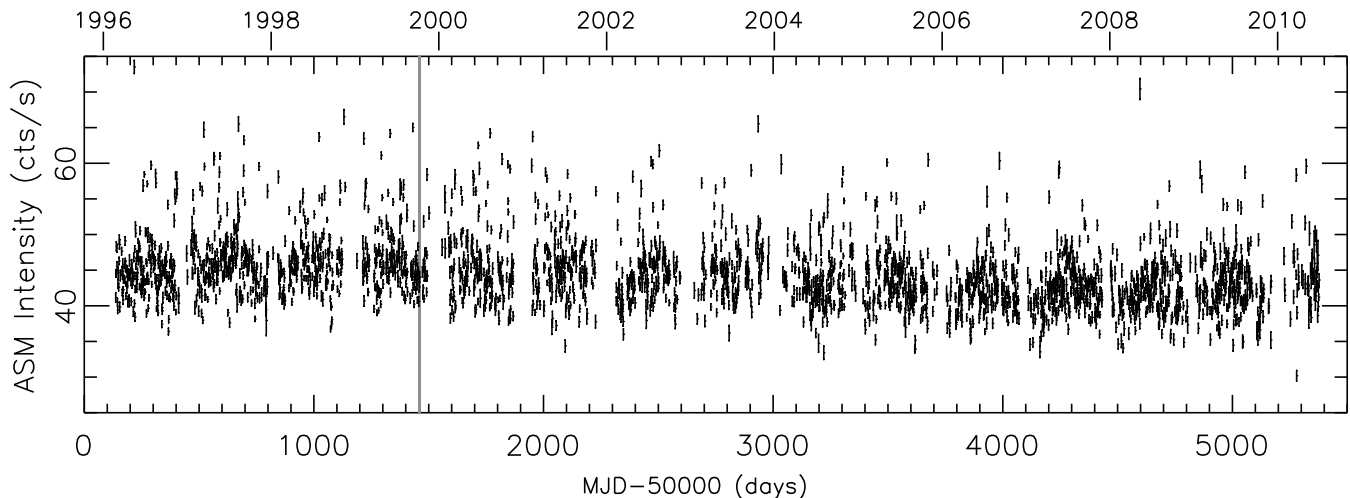} 
\caption{{\it RXTE} ASM one-day-averaged light curve of GX~17+2
  spanning $\sim$14 years.  The narrow gray region marks the interval UTC
    1999-10-03.1--12.3, during which the PCA observations analyzed in
    the paper were obtained.
\label{fig:asmlc}} 
\end{figure*}

Based on the timing and spectral properties, six of the persistently
bright neutron-star (NS) low-mass X-ray binaries (LMXBs) were
classified more than two decades ago as Z sources, named after the
spectral evolution patterns they trace out in X-ray color-color
diagrams (CDs) or hardness-intensity diagrams
\citep[HIDs;][]{hava,va2006}. These sources are \object{Sco X-1},
\object{GX 17+2}, \object{GX 349+2}, \object{GX 340+0}, \object{GX
  5-1}, and \object{Cyg X-2}. The upper, diagonal and lower branches
of their Z-shaped tracks are called horizontal, normal, and flaring
branches (HB/NB/FB), respectively. Based on the shape and orientation
of the Z tracks, these Z sources were further divided into two
subgroups, with the first three called ``Sco-like'' Z sources and the
latter three called ``Cyg-like'' \citep{kuvaoo1994}. The Sco-like Z
sources have a more vertically oriented HB and a stronger FB than the
Cyg-like types. The Z tracks themselves can also move and change
shapes in the CDs/HIDs (secular changes), most substantially in the
case of \object{Cyg X-2}.

Recently, \object{XTE J1701-462} became the first X-ray transient
identified as a Z-source, and studies of this source have
significantly improved our understanding of the spectral evolution in
Z sources and their relation to atoll sources, another subclass of NS
LMXBs with lower X-ray luminosity ($L_{\rm X}$) than Z sources.
\object{XTE J1701-462} experienced a long outburst in 2006-2007, and
it showed successive characteristics of Cyg-like Z, Sco-like Z and
atoll sources during the decay of the outburst \citep[][LRH09
  hereafter]{hovawi2007,hovafr2010,lireho2009}. During these secular
changes in the Z tracks, the upper (HB/NB) and lower (NB/FB) vertices
each evolved along a specific line in the HID (LRH09). All three Z
branches were present when the source was bright. During the decay,
the HB, NB, and FB successively disappeared, and the lower vertex
finally transitioned into the atoll soft state.

Using an X-ray spectral model that was successfully applied to two
atoll transients \citep{lireho2007}, LRH09 showed that it is most
likely the mass accretion rate ($\dot{M}$) into the disk that drives
the secular changes of Z tracks and the transitions from Z to atoll
types. While the inner disk radius remained constant in the soft state
of the atoll stage from the same spectral model, the inner disk radius
in the Z stage increased with luminosity, which was interpreted as an
effect of the local Eddington limit. On the other hand, the motion
along a Z branch on short timescales appeared to operate at roughly
constant $\dot{M}$, at least for Sco-like Z tracks
(LRH09). Furthermore, the three Z branches were linked to different
mechanisms by LRH09. The source ascends the HB as Comptonization of
the disk emission increases. The apparent luminosity of the boundary
layer increases along the NB from the lower to the upper vertices,
which can be explained as a transition from a geometrically thin disk
to a thick disk that is expected to admit an advective component of
mass flow to the NS. The FB is traced out as the inner disk radius
temporarily decreases toward the value seen in the atoll soft state,
presumably the innermost stable circular orbit (ISCO).

In this work we study \object{GX 17+2}. Figure~\ref{fig:asmlc} plots
its long-term light curve from the \xte\ All-Sky Monitor
\citep[ASM;][]{lebrcu1996}. The source shows a roughly constant level
($\sim$45 counts/s), with frequent flares on top, suggesting little
change of the overall properties of this source over a long timescale.
In fact, it shows very small secular changes
\citep{wihova1997,hovajo2002}, in contrast to \object{XTE J1701-462}.

X-ray Spectral studies of \object{GX 17+2} have been reported
previously \citep{distro2000, fafrza2005, mimife2007}.
\citet{distro2000} fitted the {\it BeppoSAX} spectra from the HB and
NB with a single-temperature blackbody (BB) plus a Comptonized
component for the continuum spectra. They found a hard tail on the HB,
which was fit by a power law (PL) and contributed $\sim$$8\%$ of the
0.1--200 keV source flux. This component gradually faded as the source
moved toward the NB, where it was no longer detectable \citep[see
  also][]{fafrza2005}. \citet{mimife2007} carried out simultaneous
radio and X-ray observations of \gx\ and found that a positive
correlation between the radio flux density and the flux of the hard
tail. Detailed timing analyses of \gx\ were made by
\citet{kuvaoo1997,wihova1997,hovajo2002}, with the evolution of
various types of quasi-periodic oscillations (QPOs) along the Z track
obtained.

Here we carry out the spectral modeling of \object{GX 17+2}, in order
to determine whether its spectral evolution is similar to that of the
Sco-like Z stage in \object{XTE J1701-462} (LRH09). As its secular
changes are small, we just concentrate on a 9-day period of intensive
\xte\ pointed observations in 1999. Different from previous X-ray
spectral modeling of this source, we use a low-Comptonization model
dominated by a thermal disk and a thermal boundary layer component,
because of its successes with atoll sources and \object{XTE J1701-462}
\citep[][LRH09]{lireho2007,lireho2010}. One main advantage of this
model over other commonly used models \citep[see][for a
  review]{ba2001} is that it infers the disk in the soft state of
atoll sources to behave approximately as $L\propto T^4$, which is
often seen in the thermal state of black hole X-ray binaries as well
\citep{lireho2007}. We also compare our spectral fit results with the
evolution of kHz QPOs and HB oscillations (HBOs) in order to obtain
hints on the origins of these QPOs. \object{GX 17+2} is more suitable
than \object{XTE J1701-462} for such a study, since such QPOs were not
detected as frequently in the latter source
\citep{hovafr2010,sameal2010,babami2011}. In
Section~\ref{sec:reduction}, we describe the reduction of the data and
the procedure by which we create our spectra. The CDs and HIDs are
also presented in this section. We describe the spectral models in
Section~\ref{sec:spectralfit}. The spectral fit results and
correlations with fast varibility are given in Section~\ref{sec:res}.
In Section~\ref{sec:dis}, we discuss the mass accretion rate, the
physical interpretation of spectral evolution, and the possible
origins of kHz QPOs and HBOs. Finally we present our conclusions.

\section{DATA ANALYSES AND COLOR-COLOR DIAGRAMS}

\begin{figure*}
\centering
\includegraphics[width=0.65\textwidth]{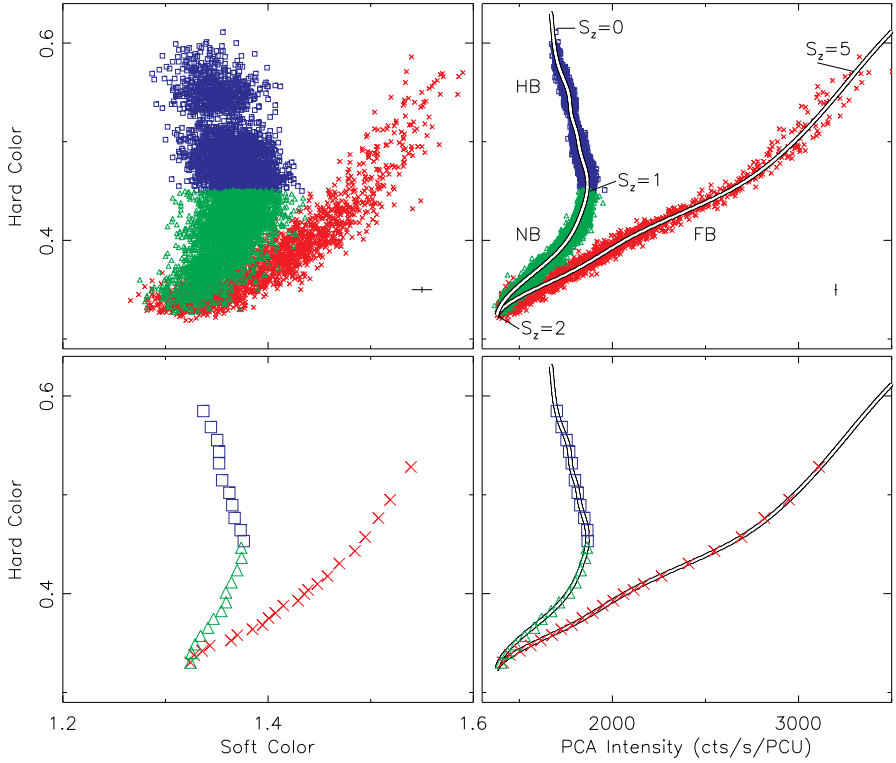} 
\caption{The color-color and hardness-intensity diagrams. The HB, NB
  and FB are marked by blue squares, green triangles, red crosses,
  respectively. Upper panels: 32 s spectra are used, and the splines
  that are used for the $S_{\rm Z}$ parametrization are shown in the
  HID. Lower panels: $S_{\rm Z}$-resolved spectra are used, and the
  splines are repeated here for reference.
\label{fig:cd}} 
\end{figure*}

\begin{figure*}
\centering
\includegraphics[width=0.70\textwidth]{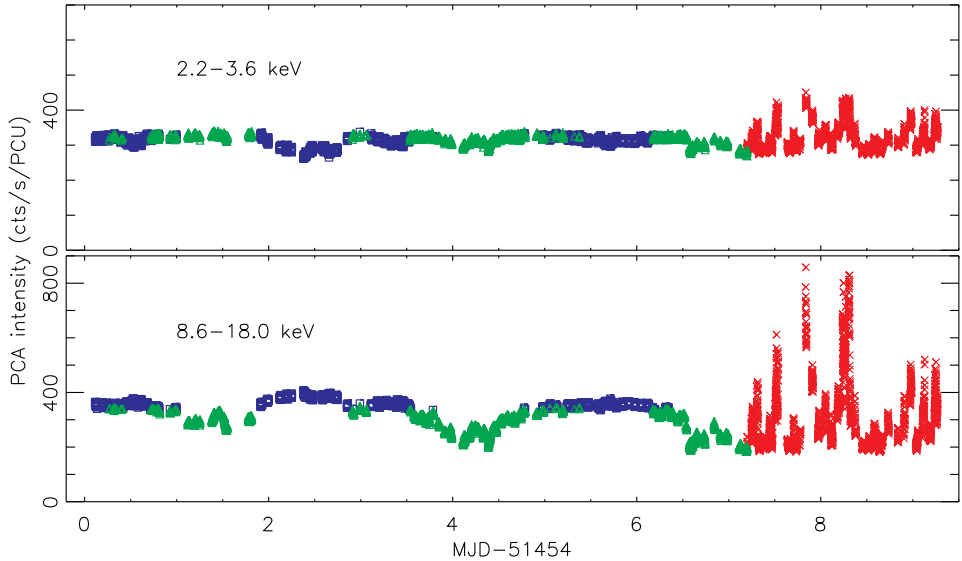} 
\caption{\xte\ PCA 32-s light curves in two energy bands. The typical
error bars are smaller than the symbol size. The meanings of the
symbols are the same as Figure~\ref{fig:cd}.
\label{fig:lc}} 
\end{figure*}

\label{sec:reduction}
We analyzed 68 \xte\ observations of \gx\ made between UTC 1999
October 3.1--12.3 (i.e., MJD 51454.1--51463.3), using the same
standard criteria to filter the data (e.g., removal of five type I
X-ray bursts) as described in \citet{lireho2007}. FTOOLS 6.9 was
used. The source can move on timescales as short as minutes in the
CD/HID, especially on the FB. To track the evolution along the Z track
but also to gain enough statistics for spectral modeling and timing
analysis, data are normally split into short exposures and then
rebinned based on their positions in the CD/HID \citep[e.g.,
  LRH09;][]{hovajo2002}. We also followed this procedure here. We
calculated X-ray colors using the Proportional Counter Unit (PCU) 2 of
the Proportional Counter Array \citep[PCA;][]{jaswgi1996}, in the same
way as described in \citet{lireho2007}. The soft color (SC) and the
hard color (HC) are defined as the ratios of the
Crab-Nebula-normalized count rates in the (3.6--5.0)/(2.2--3.6) keV
bands and the (8.6--18.0)/(5.0--8.6) keV bands, respectively. We used
spectra with exposures of 32~s from the ``standard 2'' data to
calculate the colors and construct the CD and HID, which are shown in
the upper panels of Figure~\ref{fig:cd}. The HB, NB, and FB are
denoted by the blue square, green triangle, and red cross symbols,
respectively, and this convention is used for all other figures in
this paper.

Only a single Z track is seen in these diagrams. The lower parts of
the NB and FB have substantial overlap. From the 2.2--3.6 keV and
8.6--18.0 keV light curves in Figure~\ref{fig:lc}, we see that the
source experienced frequent flares between MJD 51461.2 and 51463.3. We
identified the data within this interval as the FB, as supported by
their positions in the CD and HID in Figure~\ref{fig:cd}. The
transition from the HB to the NB is quite smooth. Their boundary is
defined to be ${\rm HC}\simeq 0.45$ (strictly $S_{\rm Z}=1$, see
below). We note that the branches are hard to separate completely and
that the above identification is approximate. 

The data points are often assigned a rank number $S_{\rm Z}$ to track
their positions along the Z track \citep{havaeb1990,hevawo1992}. This
can be done by creating splines along the Z track. Considering that
the scattering is large in the CD (mostly due to large statistical
uncertainties of SCs), we created the splines in the HID, which are
shown in the top right panel of Figure~\ref{fig:cd}. Due to large
overlap of lower parts of the FB and the NB, we created a spline for
the FB (the FB spline) and another one for the HB and NB (the HB+NB
spline). We picked the normal points of the splines by hand, with one
of them shared by the two splines. The $S_{\rm Z}$ value of this
common normal point is set to be 2.0 (the lower vertex), while the
point with HC=0.45 in the HB+NB spline has $S_{\rm Z}=1.0$ (defined as
the upper vertex and the boundary the HB and NB). $S_{\rm Z}$ values
at other positions in the splines are determined based on their
distances to these two points along the splines. Considering the
different units of HC and intensity, we divide them by a
characteristic number before calculating the distances, 0.45 for the
HC and 2000 cts/s/PCU for the intensity. Finally, the values of
$S_{\rm Z}$ for data points on the FB and on the HB and NB are
obtained by projecting them onto the FB and the HB+NB splines,
respectively. We note that in \citet{hovajo2002} a single spline was
used and was applied in the CD. This is due to wider energy bands that
they used to define the SCs. We chose to keep our definition of colors
as adopted in \citet{lireho2007} and LRH09 for easy comparison with
these studies.

We then combined 32 s spectra from PCU 2 based on selections in
$S_{\rm Z}$ to create spectra with longer exposures for our spectral
fits. The ranges of $S_{\rm Z}$ and the exposures for these $S_{\rm
  Z}$-resolved spectra are given in
Tables~\ref{tbl-fitparmod1}--\ref{tbl-fitparmod2}. Their locations in
the CD and HID are shown in the lower panels of
Figure~\ref{fig:cd}. We also created spectra of PCU 0 and those of the
High Energy X-ray Timing Experiment \citep[HEXTE;][]{roblgr1998},
which match the spectra of PCU 2 in time. PCUs 0 and 2 were the only
units that were operating during all of our observations. For the
HEXTE, we used both Clusters A and B. We applied dead time corrections
to both the PCA and HEXTE data.

We also created power density spectra (PDS), which match the $S_{\rm
  Z}$-resolved spectra in time, to search for the kHz QPOs and HBOs
and study their relations with our spectral fit parameters. As we used
a different way of data reduction from \citet{hovajo2002}, our PDS do
not match theirs in time. We used the PCA event files to create 5.8--40
keV light curves (from all available PCUs) in a time bin of $2^{-12}$
s. We then calculated $S_{\rm Z}$-resolved PDS by averaging the
individual PDS computed for the appropriate set of 32-s time
intervals. We excluded events below 5.8 keV to increase the detection
significance of QPOs, as in \citet{hovajo2002}. We refer to their
paper for more information on the data modes of the event files. We
then fitted the PDS in a very similar way as \citet{hovajo2002} to
search for kHz QPOs and HBOs. Basically the 200-2048 Hz PDS were
fitted with one or two Lorentzians for the kHz QPOs plus a
dead-time-modified Poisson level, and the 0.03125--256 Hz PDS were
fitted with a PL, a cutoff PL (CPL) and/or a Lorentzian for various
broad noise components and Lorentzians for QPOs, including HBOs, plus
a constant for the Poisson level. We also calculated the rms within
the 0.1--10 Hz frequency band from these PDS, in order to study its
correlation with the Comptonization fraction inferred from the
spectral parameters.

As the PCA does not cover the low energy band below 2.5 keV and has
only a modest energy resolution, we also analyzed two {\it Suzaku}
observations (sequence numbers 402050010 and 402050020), using the
same data reduction procedure as in \citet{lireho2010}, to help to
estimate the absorption and the Fe line energy for our spectral fits
to the {\it RXTE} data. The calibration files of 2010 June and {\it
  Suzaku} FTOOLS version 15 were used. Spectra from XIS
\citep{kotsdo2007} 0 and 3 were extracted, with the central region of
a radius of 55 pixels excluded to reduce the event pileup effect. XIS
1 was not used as it was in the full window mode, resulting in serious
event pileup. We also used the spectra of the PIN diodes of the Hard
X-ray Detector \citep{taaben2007}.

\section{SPECTRAL MODELING}
\label{sec:spectralfit}

\begin{figure}
\centering
\includegraphics{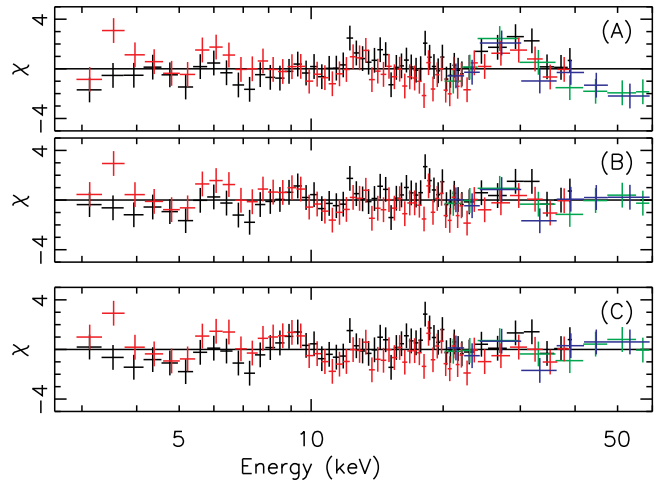}
\caption{The fit residuls in terms of sigma for the spectrum with
  $S_{\rm Z}$=0.32--0.40, using various models for the Comptonization
  (see text for details). This spectrum is used considering its large
  Comptonization fraction and exposure. (A): a PL, $\chi^2_\nu$=1.13;
  (B) a CPL, i.e., Model~1, $\chi^2_\nu$=0.80; (C) an nthComp, i.e.,
  Model~2, $\chi^2_\nu$=0.86. The black, red, green, and blue data
  points are for the PCU2, PCU0, HEXTE Clusters A and B, respectively.
\label{fig:comdelchi}} 
\end{figure}

\begin{figure*} \epsscale{1.0}
\centering
\subfigure[Model~1: MCD+BB+CPL]{
\includegraphics[scale=0.8]{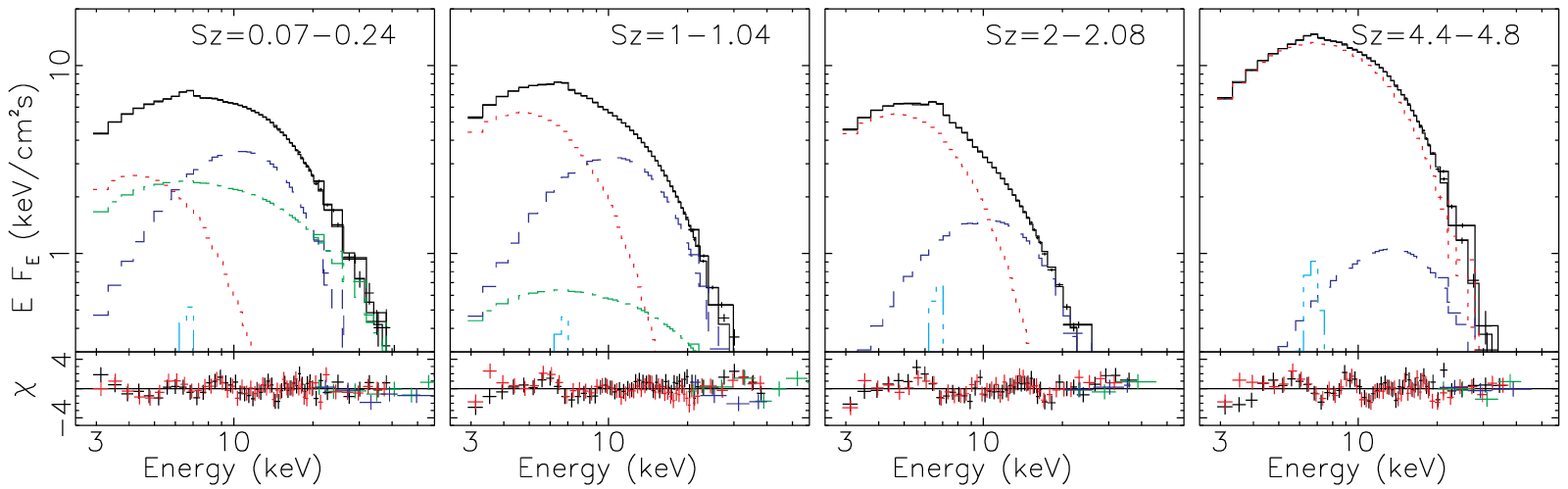}
\label{fig:ufspredm1}
}
\subfigure[Model~2: MCD+BB+nthComp]{
\includegraphics[scale=0.8]{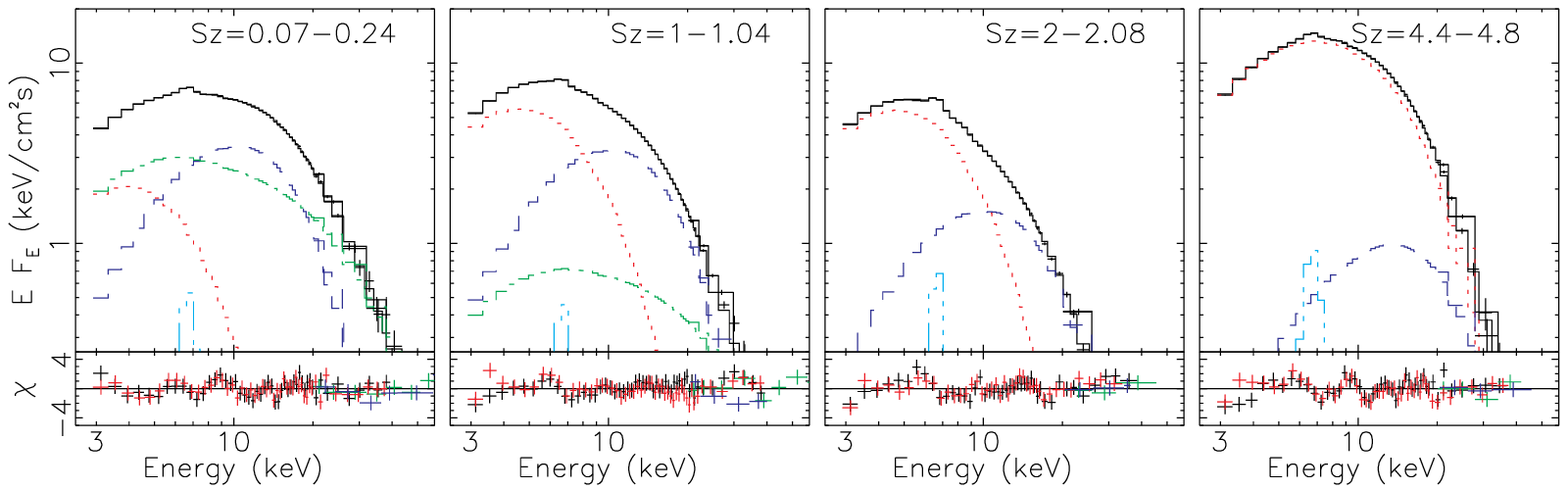}
\label{fig:ufspredm2}
}
\caption{The unfolded spectra and fit residuals for Models 1 and 2 at
  different positions in the Z track. For the unfolded spectra, only
  spectra of PCU2 and HEXTE Cluster A are shown for clarity. The total
  model fit is shown as a black solid line, the (unscattered) MCD
  component as a red dotted line, the BB component as a blue dashed
  line, the Comptonized component (CPL/nthComp) as a green dot-dashed
  line, and the Fe line as a cyan triple-dot-dashed line. For the fit
  residuals, the black, red, green, and blue data points are for the
  PCU2, PCU0, HEXTE Cluster A and B,
  respectively. \label{fig:ufspred}}
\end{figure*}
 
The $S_{\rm Z}$-resolved PCA (PCU0 and PCU2) and HEXTE (Clusters A and
B) spectra were fitted jointly, with their relative normalizations
allowed to float. We used the energy range 2.9--40.0 keV for the PCA
spectra and 20.0--60.0 keV for the HEXTE spectra. We assumed a model
systematic error of 0.5\%, as recommended by the PCA team. Both PCA
and HEXTE data were binned to have at least 40 counts per bin. All
error bars of spectral fit results quoted are at a 90\%-confidence
level, unless indicated otherwise.

We fitted the spectra with a model consisting of a BB ({\it bbodyrad}
in XSPEC), used to describe the boundary layer, a multicolor disk
blackbody (MCD; {\it diskbb} in XSPEC), and a weak Comptonized
component. It also includes a Fe K Gaussian line, a Fe K absorption
edge, and an interstellar medium absorption, described by the {\it
  gaussian}, {\it edge}, and {\it wabs} models in XSPEC,
respectively. The absorption edge was included in \citet{distro2000}
in their fits to the {\it BeppoSAX} spectra, and we found that
including it in our model also improved our fits significantly
($\chi^2$ decreases by about 37 on average for each $S_{\rm
  Z}$-resolved spectrum, with 130 degrees of freedom typically). To
reduce the scattering of our results, we used fixed values of the
hydrogen column density $N_{\rm H}=2.35\times$$10^{22}$ cm$^{-2}$, the
Gaussian line central energy $E_{\rm ga}=6.64$ keV, and the edge
energy $E_{\rm edge}=8.93$ keV (derived below and at the end of this
section) for all fits. We explored several descriptions of the
Comptonized component. It is significant on the HB, and understanding
this component is important toward our understanding of the HB, the
HBOs, and the kHz QPOs.

The use of three spectral components to fit X-ray spectra of accreting
NSs in the soft state must be done carefully
\citep[see][]{lireho2007}. The Comptonized and the thermal components
can become degenerate when the Comptonized component is steep (e.g.,
the photon index larger than 2.5 when the Comptonization is modeled as
a PL). Here, we first used a CPL ({\it cutoffpl} in XSPEC) to describe
the Comptonized component. The whole model is
$wabs(diskbb+bbodyrad+{\it cutoffpl}+gaussian)edge$ using the XSPEC
terminology, and this is our first model (Model~1, or model
MCD+BB+CPL, hereafter). Using this model, the Comptonized component
turned out to be weak compared with the thermal components except at
the top of the HB. The photon index $\Gamma_{\rm CPL}$ and the
e-folding energy $E_{\rm CPL}$ were generally not well constrained
from the fits, and they showed no clear sign of variation along the Z
track. To have a better constrained estimate of them, we fitted all
the HB spectra simultaneously with these two parameters tied to be the
same and obtained $\Gamma_{\rm CPL}=1.40$$\pm$0.14 and $E_{\rm
  CPL}=9.9$$\pm$1.0 keV. These values were fixed in the final fits to
all spectra. We then obtained $E_{\rm edge}=8.93\pm0.03$ keV from the
simultaneous fit to the spectra from all branches. Although fits with
the Comptonized component described by a PL can also have the
$\chi^2_\nu$ values around one, such fits have systematic residuals
above 20 keV on the HB (Figure~\ref{fig:comdelchi}), which are
significantly reduced using a CPL. The $\chi^2$ decrease is 51.7, from
173.5 (d.o.f=153) to 121.8, for the $S_{\rm Z}$=0.32--0.40
spectrum. Applying the posterior predictive $p$-value method
\citep{huvaos2008,prvaco2002} to this spectrum, we found that the
reduction of $\chi^2$ from the introduction of $E_{\rm CPL}$ is less
than 15 for all $10^4$ spectra that we simulated. Thus the exponential
rollover is required for the Comptonized component at a confidence
level above 99.99\% (at least for the $S_{\rm Z}$=0.32--0.40
spectrum).  Figure~\ref{fig:ufspredm1} shows the unfolded spectra and
fit residuals at four representative positions of the Z track (the top
of the HB, the upper vertex, the lower vertex, and the top of the FB)
using Model~1.

Next, we attempted to describe the Comptonization in a self-consistent
way. Considering that the CPL luminosity variation along the HB is
strongly anti-correlated with the disk luminosity from Model~1
(Section~\ref{sec:res}), we assumed the scenario that there is a
corona above the disk so that some photons from the thermal disk
emission are scattered by the hot electrons in it and turned into
Comptonization emission. Such a picture has been suggested in some
studies of black hole X-ray binaries and can be modeled with the SIMPL
model \citep[in XSPEC,][]{stnamc2009, stmcre2009}. SIMPL is an
empirical convolution model of Comptonization in which a fraction of
the photons from an input seed spectrum are scattered into a power-law
component with the rest unscattered and observed directly. While
applying the SIMPL model with the MCD as the input seed spectrum to
our spectra at the top of the HB where the Comptonization is the
strongest, we found some systematic fit residuals above 20 keV, with
the data falling below the model prediction at high energies (similar
to case A in Figure~\ref{fig:comdelchi} using a PL to describe the
Comptonization). A possible explanation for this is that there is a
high-energy cut-off in the Comptonization emission (i.e., the corona
temperature is close to or within the energy range of our data), which
was not included in the SIMPL model for simplicity
\citep{stnamc2009}. To account for this, one option is to modify the
SIMPL model to include a high-energy cut-off for the Comptonized
component by multiplying it by, e.g., the {\it highecut} model in
XSPEC. An alternative option is to replace SIMPL with a Comptonization
model with the corona temperature as a parameter, such as the nthComp
model \citep[in XSPEC,][]{zydosm1999,zdjoma1996,lizd1987}. The nthComp
model approximates the Comptonization by solving the Kompaneets
equation with a relativistic correction to energy transfer between
photons and electrons. The input seed photons can be a BB or a MCD in
spectral shape, and we assumed the latter throughout the paper. The
key parameters of this model include the asymptotic power-law photon
index $\Gamma_{\rm nthComp}$, the corona electron temperature $kT_{\rm
  e, nthComp}$, the seed photon temperature, and the normalization
$N_{\rm nthComp}$.

In this paper, we present the results with the Comptonized component
modeled with nthComp. We note that similar conclusions can also be
drawn using the SIMPL model modified to include the high-energy
cut-off. Then our second model (Model~2 or model MCD+BB+nthComp,
hereafter) is: $wabs(diskbb+bbodyrad+nthcomp+gaussian)edge$, with the
seed photon temperature tied to the temperature of the MCD component
$kT_{\rm MCD}$. With Model~2, our aim is to take into account the
photons Compton scattered so as to track the behavior of the thermal
disk emission prior to Compton scattering (denoted as MCDPS
hereafter), as done in SIMPL. Therefore, we calculated the photon flux
of the MCDPS by adding the photon flux of nthComp (the scattered part
of the MCDPS) to that of the MCD component (the unscattered part) and
then obtained the corresponding energy flux and normalization of the
MCDPS by increasing those of the MCD component proportionally to the
photon flux (the MCDPS temperature $kT_{\rm MCDPS}$ is the same as
$kT_{\rm MCD}$).

To have a better constrained estimate of the parameters of the
Comptonized component, we fitted the HB spectra simultaneously with
their $\Gamma_{\rm nthComp}$ and $kT_{\rm e, nthComp}$ tied to be the
same, as we did for Model~1. We found that comparably good fits can be
obtained using a large range of $\Gamma_{\rm nthComp}$, i.e.,
2.0--2.7, with $kT_{\rm e, nthComp}$ of 5.6--7.3 keV
correspondingly. This uncertainty is coupled to the uncertainty in the
fraction of the thermal disk emission scattered into the
Comptonization emission, with a larger scattering fraction inferred
using a larger value of $\Gamma_{\rm nthComp}$. However, the
parameters (the temperature and the normalization) of the MCDPS and
the BB are much less affected. Allowing a high value of $\Gamma_{\rm
  nthComp}$ might have the risk of large interference between the
thermal (especially the disk) and the Comptonized components, as
similarly observed in the fits to atoll-source X-ray spectra in
\citet{lireho2007}. In the end, we chose to present results using
$\Gamma_{\rm nthComp}=2.3$. Then the fit to all the HB spectra
simultaneously gave $kT_{\rm e, nthComp}=6.04\pm0.23$. The above
values of $\Gamma_{\rm nthComp}$ and $kT_{\rm e, nthComp}$ were fixed
in the final fits to all spectra. Sample fits with Model~2 are given
in Figure~\ref{fig:ufspredm2}. The systematic effect of our choice of
$\Gamma_{\rm nthComp}$ will be discussed later by comparing with
results using other values of $\Gamma_{\rm nthComp}$ in the range of
2.0--2.7.

We estimated $N_{\rm H}$ and $E_{\rm ga}$ with the two {\it Suzaku}
observations. The energy bands and the data binning method used were
the same as in \citet{lireho2010}. We applied Model~1 (Model~2 gives
very similar results) and fitted both {\it Suzaku} observations
simultaneously with $N_{\rm H}$ and $E_{\rm ga}$ tied to be the same
for both observations. Only a very weak CPL is needed in both
observations. The absorption edge cannot be constrained by the data,
probably because of the gap between the XIS and PIN (10--11 keV) and
the relatively poor data quality around this edge. We still included
it in the fit but fixed $E_{\rm edge}=8.93$ keV (see above). Finally,
we obtained $N_{\rm H}=(2.35$$\pm$$0.01)\times$$10^{22}$ cm$^{-2}$ and
$E_{\rm ga}=6.64$$\pm$$0.06$ keV. This $N_{\rm H}$ is consistent with
$N_{\rm H}=(2.38$$\pm$$0.12)\times$$10^{22}$ cm$^{-2}$ obtained by
\citet{wrguoz2008} from the modeling of absorption edges found in {\it
  Chandra} high-resolution X-ray spectra.

\section{RESULTS}
\label{sec:res}
\subsection{Spectral Evolution}
\label{sec:spev}

\begin{figure}
\centering
\includegraphics{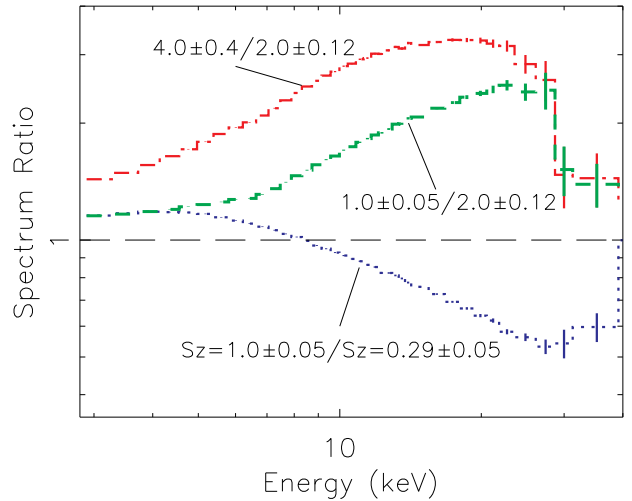}
\caption{The ratios of the spectra on the two ends of each
branch. Spectra with high total PCA intensity are divided by those
with lower total PCA intensity in order to illustrate that on each
branch the spectrum evolves differently.
\label{fig:ratio}} 
\end{figure}

\begin{figure*} \epsscale{1.0}
\subfigure[Model~1: MCD+BB+CPL]{
\includegraphics[scale=1.0]{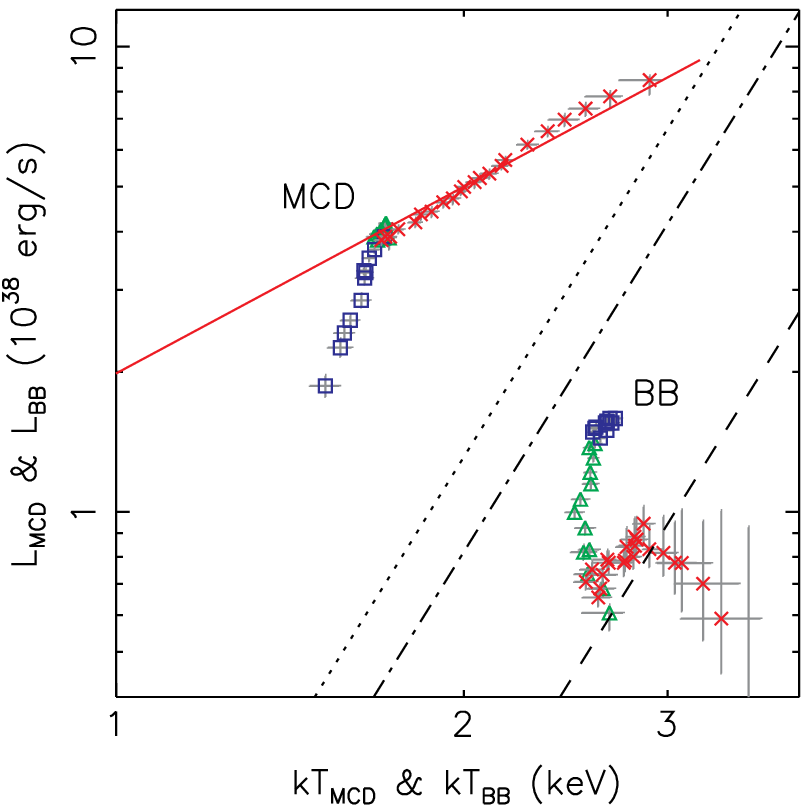}
\label{fig:Tlumm1}
}
\subfigure[Model~2: MCD+BB+nthComp]{
\includegraphics[scale=1.0]{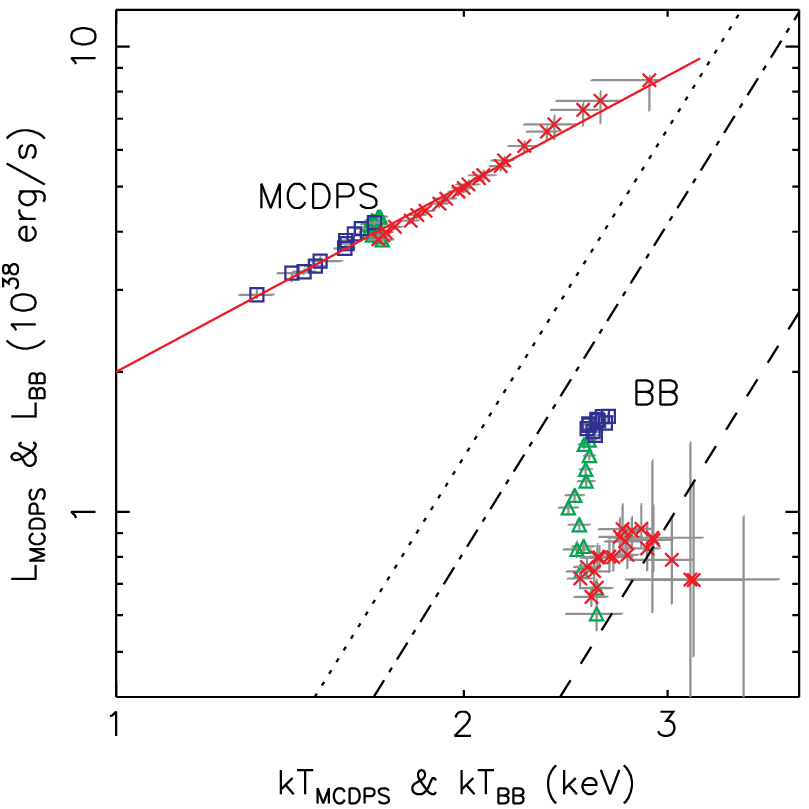}
\label{fig:Tlumm2}
}
\caption{The fit results of $S_{\rm Z}$-resolved spectra. The
  luminosities of the thermal components are plotted against their
  characteristic temperatures. The dotted, dot-dashed, and dashed lines correspond to $R=8.0$, 6.3, and 3.0 km, respectively, assuming $L_{\rm X}=4\pi R^2\sigma_{\rm SB} T^4$, and the red solid line is a constant
  $\dot{M}$ line (see text). Note that for Model~1 we plot the thermal
  disk emission unscattered and observed directly, while for Model~2
  we plot the thermal disk prior to scattering. \label{fig:Tlum}}
\end{figure*}
 %%%%\clearpage

\begin{figure*} \epsscale{1.0}
\subfigure[Model~1: MCD+BB+CPL]{
\includegraphics[scale=1.0]{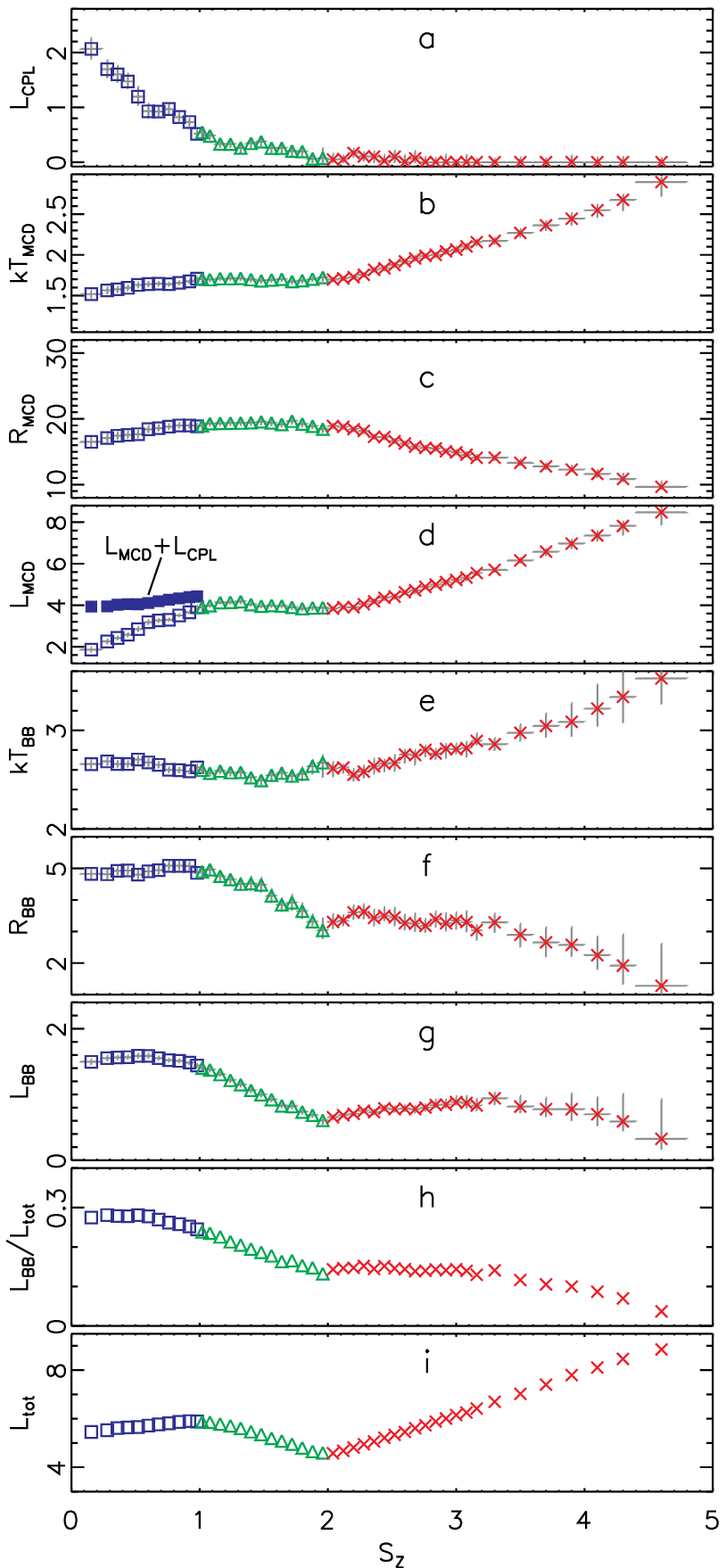}
\label{fig:parm1}
}
\subfigure[Model~2: MCD+BB+nthComp]{
\includegraphics[scale=1.0]{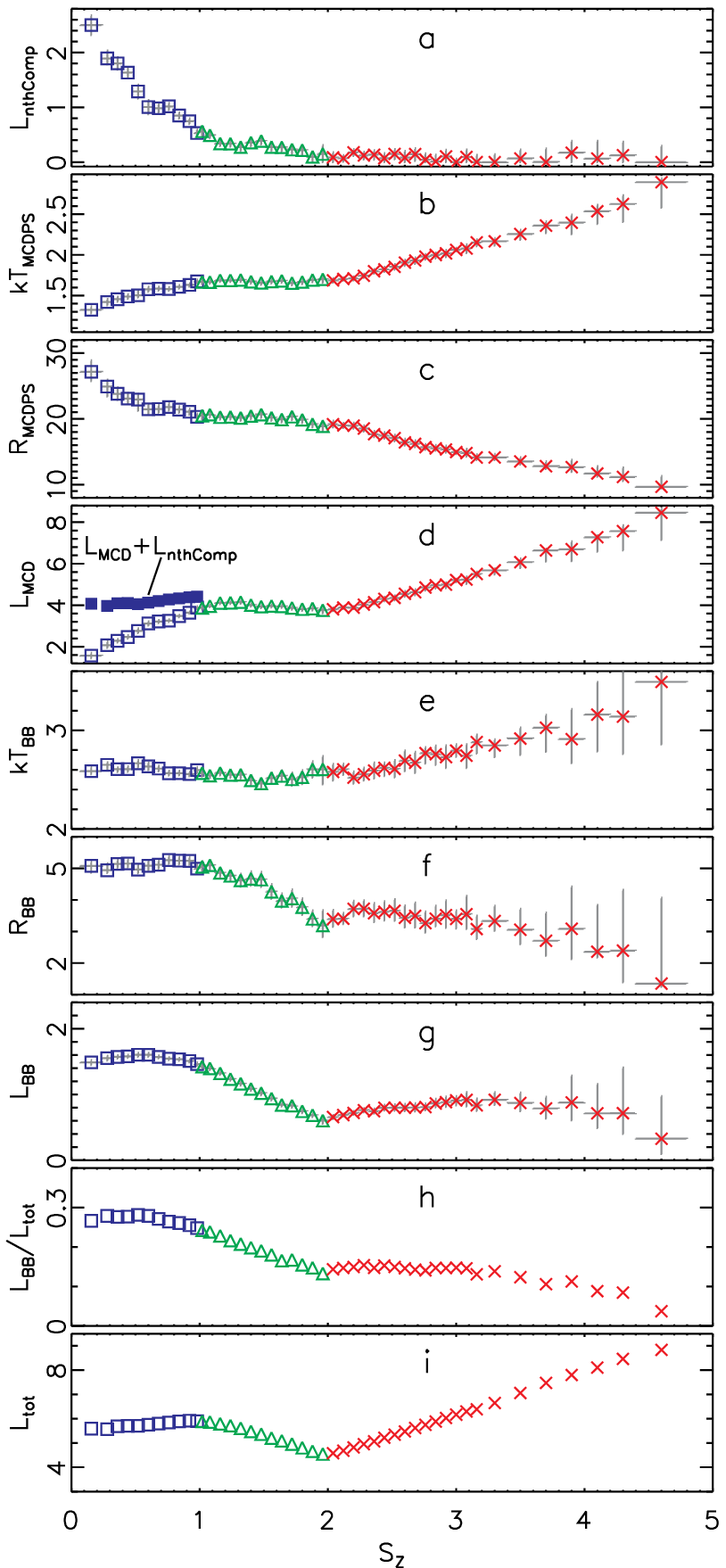}
\label{fig:parm2}
}
\caption{The fit results of $S_{\rm Z}$-resolved spectra as a function
of the rank number $S_{\rm Z}$. The error bar of $S_{\rm Z}$
corresponds to the range of $S_{\rm Z}$ for each spectrum. $L_{\rm tot}$ is the
total luminosity, whose error bar is not plotted but is given in
Tables~\ref{tbl-fitparmod1}--\ref{tbl-fitparmod2}. See text for the
meanings of other quantities. The units of the $R$, $kT$, $L$
quantities are km, eV, and $10^{38}$ erg~s$^{-1}$,
respectively. \label{fig:parameters}}
\end{figure*}

%%%%\clearpage
\begin{figure}
%\figurenum{9}  
\plotone{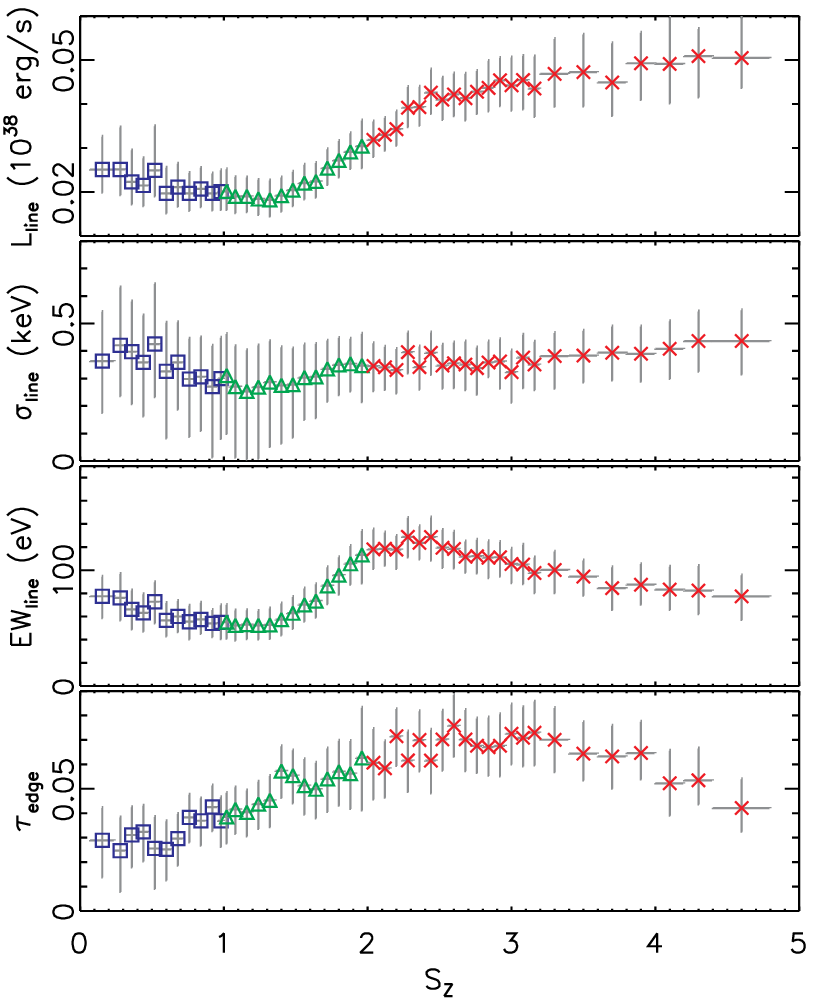}
\caption{The fit results of Gaussian Fe line as a function of the rank
  number $S_{\rm Z}$ using Model~1. \label{fig:gaparameters}}
\end{figure}
%%%%\clearpage

Figure~\ref{fig:lc} shows that the source goes back and forth between
the HB and NB several times in the first seven days and then up and
down the FB many times in the last two days, but no clear secular
changes are seen in Figure~\ref{fig:cd} (upper panels). As in LRH09,
we calculate the ratios of spectra in different branches, which are
shown in Figure~\ref{fig:ratio}. The blue dotted line is for the HB
(the upper vertex divided by the top of the HB) and shows that the
spectrum pivots around 9 keV, with the intensity below increasing and
above decreasing as the source descends the HB. The green dashed line
corresponds to the NB. Although the intensity increases over the whole
energy range shown as the source evolves up the NB, the effect becomes
stronger above $\sim$7 keV. The increase of the intensity on the FB is
also mostly in the high energies 10--30 keV (red dot-dashed line). The
ratios for the NB and FB drop sharply around 30 keV, above which in
fact there is no significant source emission. These ratios are similar
to \object{XTE J1701-462} in the Sco-like stage (LRH09).

The parameters derived from the spectral fits are shown in
Figures~\ref{fig:Tlum}--\ref{fig:gaparameters} and are listed in
Tables~\ref{tbl-fitparmod1}--\ref{tbl-fitparmod2}. For all luminosity
($L$) and radius ($R$) quantities, we assume a source distance of 12.6
kpc (Section~\ref{sec:discburst}), and the luminosities are based on
bolometric fluxes of relevant spectral components (for the CPL
component, we integrate over energies above 1 keV). Isotropic emission
is assumed for all components (but see discussion in
Section~\ref{sec:reszbranch}), except the MCD one, whose inclination
$i$ is assumed to be 60$\degr$. $R_{\rm BB}$ is the apparent radius of
the BB component and is related to the normalization as $N_{\rm
  BB}\equiv (R_{\rm BB, km}/D_{\rm 10 kpc})^2$, where $D_{\rm 10 kpc}$
is the distance to the source in units of 10 kpc. $R_{\rm MCD}$ in
Model~1 is the apparent inner disk radius and is calculated from the
MCD normalization $N_{\rm MCD}\equiv (R_{\rm MCD, km}/D_{\rm 10
  kpc})^2\cos i$, while $R_{\rm MCDPS}$ in Model~2 is from the MCDPS
normalization $N_{\rm MCDPS}\equiv (R_{\rm MCDPS, km}/D_{\rm 10
  kpc})^2\cos i$.

Figure~\ref{fig:Tlum} shows the luminosities of the thermal components
MCD/MCDPS and BB versus their characteristic temperatures, while
Figures~\ref{fig:parameters} and \ref{fig:gaparameters} show the
spectral fit results as a function of the rank number $S_{\rm Z}$. We
included several constant radius lines in Figure~\ref{fig:Tlum},
assuming $L_{\rm X}=4\pi\sigma_{\rm SB}R^2T^4$. The dotted and
dot-dashed lines have $R=8$ km ($N_{\rm BB}=40$) and $R=6.3$ km
($N_{\rm BB}=25$), corresponding to the apparent sizes of the NS from
the sum of the boundary layer and the burst emission and from only the
burst emission, respectively (Section~\ref{sec:discburst}). The dashed
lines have $R=3$ km, shown for reference. The red solid lines describe
the relation between $L_{\rm MCD}$ and $kT_{\rm MCD}$ with varying
$R_{\rm MCD}$ at a constant $\dot{M}$, i.e., $L_{\rm MCD} \propto
T_{\rm MCD}^{4/3}$ (LRH09; referring to the MCDPS in Model~2).

We focus on the results of Model~1 for the HB first.
Figures~\ref{fig:Tlum}--\ref{fig:parameters} (left panels) show that
$L_{\rm MCD}$ decreases and $L_{\rm CPL}$ increases as the source
climbs up the HB, while $L_{\rm BB}$ changes much less. The
anti-correlation between $L_{\rm MCD}$ and $L_{\rm CPL}$ is clear, as
their sum (filled blue squares in the left panel d of
Figure~\ref{fig:parameters}) changes by $\lesssim$10$\%$ on the HB,
much less than either individual component.

Model~2 assumes that some corona surrounds the thermal disk and
produces the Comptonization emission. From this model, we also see the
strong anti-correlation between the (unscattered) MCD component and
the Comptonization emission (Figure~\ref{fig:parameters}). However,
unlike Model~1, we see that the apparent inner disk radius from
Model~2 (inferred from the MCDPS) increases as the source climbs up
the HB. Moreover, the thermal disk based on the MCDPS follows a
constant $\dot{M}$ line (the red solid line in Figure~\ref{fig:Tlum}).

The Comptonized component becomes weak on the NB and is almost
negligible on the FB (Figures~\ref{fig:ufspred} and
\ref{fig:parameters}), and the results of other components become very
similar between Models~1 and 2 in these two branches. In
Figures~\ref{fig:Tlum}--\ref{fig:parameters}, both models show that in
the NB the largest changes occur in the BB component. Its
normalization varies by a factor of about 2.5, with the temperature
hovering around 2.6 keV
(Tables~\ref{tbl-fitparmod1}--\ref{tbl-fitparmod2}). This suggests
that the change of intensity above $\sim$7 keV (green line in
Figure~\ref{fig:ratio}) is mostly due to the BB component. However,
since the BB is just a small component ($<$30\% in terms of
luminosity), the total luminosity only changes by $\sim$20$\%$ on the
NB (Figure~\ref{fig:parameters} and
Tables~\ref{tbl-fitparmod1}--\ref{tbl-fitparmod2}).

The variations of the MCD/MCDPS component on the FB approximately
follow a constant $\dot{M}$ line (the red solid line in
Figure~\ref{fig:Tlum}) from both models. The disk temperature
increases from $\sim$1.7 keV at the lower vertex to nearly 3 keV at
the top of the FB. Compared with the dotted line for the size of NS,
the disk is consistent with being truncated outside the NS. The
results of the BB component show large uncertainties, especially in
the upper part of the FB, making it difficult to infer the evolution
trend of this component in this branch. This was also seen in the
spectral fits of \object{XTE J1701-462} in LRH09.

We have seen that the MCDPS from Model~2 evolves approximately along a
constant $\dot{M}$ line over the whole Z track. The fractional rms of
$\dot{M}$ using the relation of $\dot{M}\propto L_{\rm MCDPS}R_{\rm
  MCDPS}$ is about 6\%. Assuming $L_{\rm MCDPS}=kT_{\rm MCDPS}^\beta$,
we estimate the slope $\beta$ by minimizing the $\chi^2$ value defined
as $\sum (y_i-\beta
x_i-\alpha)^2/(\epsilon_{yi}^2+\beta^2(\epsilon_{xi}^2+\epsilon_{0}^2))$
\citep[e.g.,][]{trgebe2002}, where $y=\log(L_{\rm MCDPS})$ with
1$\sigma$ error $\epsilon_y$, $x=\log(T_{\rm MCDPS})$ with 1$\sigma$
error $\epsilon_x$, $\alpha=\log(k)$, and the subscript $i$ denotes
the data points used. $\epsilon_0$ is introduced to account for the
possible systematic error $r$ of $kT_{\rm MCDPS}$
($\epsilon_0=\log(1+r)$) and is set to such a value that the reduced
$\chi^2$ is one. We obtain $\beta=1.36\pm0.07$ from data of the whole
Z track, with $r=2.4\%$ used. Consistent values of $\beta$ can also be
obtained using other choices of $\Gamma_{\rm nthComp}$ in the range of
2.0--2.7 in Model~2.

Figure~\ref{fig:gaparameters} shows the variations of the Fe K line
and absorption edge using Model~1, while Model~2 gives very similar
results. The line luminosity $L_{\rm line}$ first slightly decreases
from the top of the HB to the upper vertex and then increases again
until the end of the FB. It is highest on the FB. The line width
$\sigma_{\rm line}$ is $\sim$0.4 keV. The equivalent width EW$_{\rm
  line}$ peaks near the lower vertex, with a value of $\sim$120 eV,
and is $\sim$60 eV around the upper vertex. The absorption depth at
the edge energy $\tau_{\rm edge}$ is around 0.05 and peaks at the
lower part of the FB. We note that as \xte\ data have only a modest
energy resolution ($\sim$1.2 keV FWHM at 6.5 keV) the results of the
Fe K line and absorption edge should be regarded with caution.

\subsection{Comparison with Fast Variability}
\label{sec:qpo}
\begin{figure}
%\figurenum{9}  
\plotone{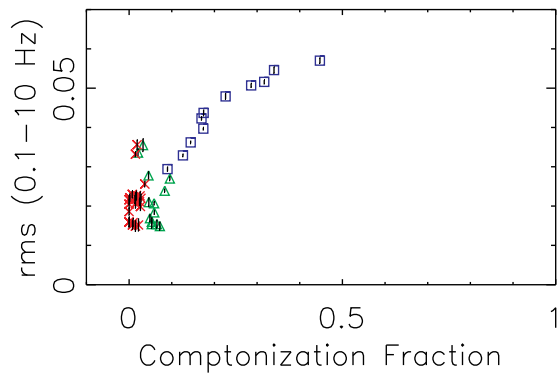}
\caption{The fractional rms versus the luminosity fraction of the Comptonized component, i.e., nthComp, obtained with Model~2.
 \label{fig:rms}}
\end{figure}

\begin{figure}
%\figurenum{9}  
\plotone{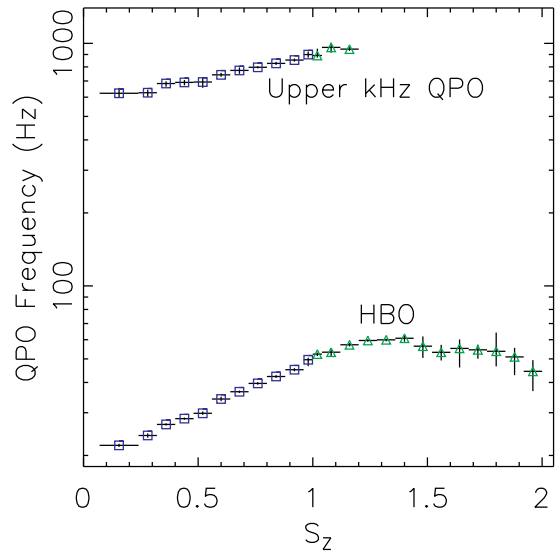}
\caption{The dependence of the frequencies of the upper kHz QPO and the HBO on $S_{\rm Z}$.
 \label{fig:fresz}}
\end{figure}

%%%%\clearpage

\begin{figure}
%\figurenum{9}  
\plotone{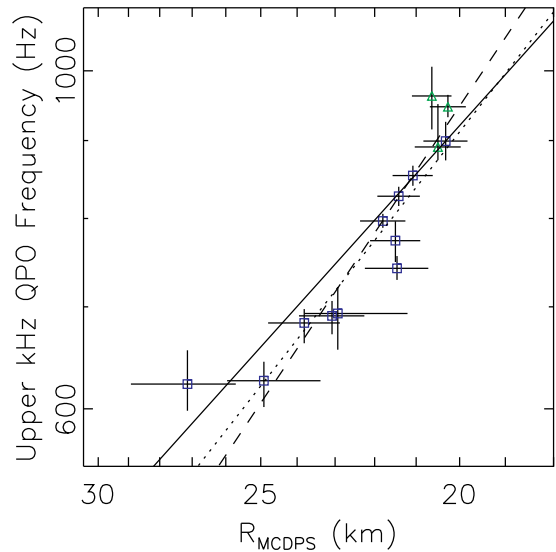}
\caption{The frequency of the upper kHz QPO versus the inner disk
  radius (prior to Compton scattering) from Model~2. The inner disk
  radius is derived from the MCDPS normalization, without corrections
  for factors such as the hardening effect. The solid, dotted, and
  dashed lines show the relation of frequency~$\propto$~$R_{\rm
    MCDPS}^{\beta}$, with $\beta=-1.5$ (corresponding to the Keplerian
  frequency at the inner disk radius), $-1.72$ (from the fit to the HB
  data), and $-2.01$ (from the fit to the HB and NB data),
  respectively.
 \label{fig:uqporin}}
\end{figure}
%%%%\clearpage

\begin{figure}
%\figurenum{9}  
\plotone{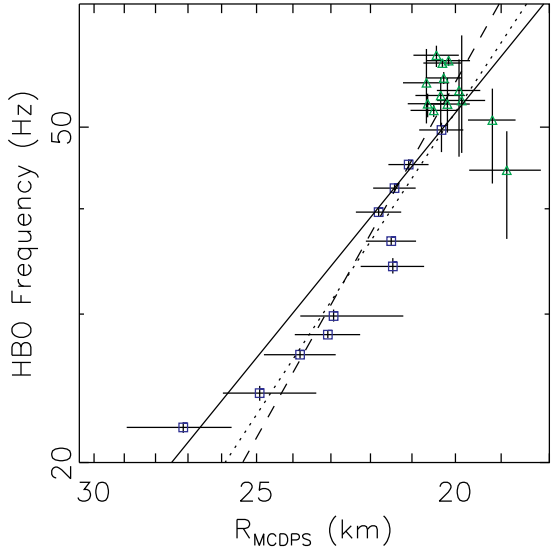}
\caption{The frequency of the HBO versus the apparent inner disk
  radius (prior to Compton scattering) from Model~2. The solid,
  dotted, and dashed lines show the relation of
  frequency~$\propto$~$R_{\rm MCDPS}^{\beta}$, with $\beta=-3$
  (corresponding to Lense-Thirring precession), $-3.70$ (from the fit
  to the HB data), and $-4.35$ (from the fit to the HB and NB data),
  respectively.
 \label{fig:hborin}}
\end{figure}
%%%%\clearpage

Figure~\ref{fig:rms} shows the rms integrated from the PDS (0.1-10 Hz)
versus the Comptonization fraction inferred from Model~2. The same
type of plot, substituting results from Model~1, appears very
similar. Both models infer low Comptonization fraction ($<$10\%) in
the NB and FB, with significantly higher values for the HB. The rms
variations appear to be roughly correlated with the Comptonization
fraction, and only the HB has rms $>$5\%. In terms of these
properties, the NB and FB are similar to the soft/thermal state of
black hole X-ray binaries and atoll sources \citep{lireho2007}, while
the HB is closer to their transitional/steep power-law state. We note
that the rms shows an enhancement near the lower vertex, which is due
to the normal branch oscillations. We also note that the FB in fact
shows large variability in light curves. The low rms obtained above to
some extent is because we only integrate the rms above 0.1 Hz. Using a
lower integration limit could result in a larger rms, as the PDS in
the FB is steep, with a power-law slope around 2
\citep{hovajo2002}. In the above, we just use the same frequency range
as in \citet{lireho2007} and LRH09 for fair comparison with their
results.

Table~\ref{tbl-upqpo} lists the HBOs (fundamental) and upper kHz QPOs
that we detected in our data set. The dependence of their frequencies
on $S_{\rm Z}$ is plotted in Figure~\ref{fig:fresz}. There are also
other QPOs such as HBO second harmonics and lower kHz QPOs, which are
not explored here due to their smaller ranges spanned in the Z track
\citep{hovajo2002}. The HBOs span the whole HB and NB ranging from
about 20 to 60 Hz, while the upper kHz QPOs span the whole HB and a
small upper part of the NB ranging from about 620 to 1000 Hz,
generally consistent with the results in \citet{hovajo2002}. However,
they found some upper kHz QPOs from around 1000 up to 1090 Hz on the
middle NB, which we could hardly detect from our data. This is
probably because they used more observations in their data analysis
plus the fact that these QPOs are weak to begin with (rms $<$
2\%). The HBOs become broader on the lower NB, and their FWHM values
have large uncertainties. Thus we fixed the FWHM to be 15 Hz
\citep{hovajo2002} in some fits.

Figure~\ref{fig:uqporin} shows the frequency of the upper kHz QPO
versus $R_{\rm MCDPS}$ from Model~2. Assuming a relation of
frequency~$=$~$kR_{\rm MCDPS}^{\beta}$ and using the slope estimate
method in Section~\ref{sec:spev}, we obtain
$\beta=-1.72^{+0.32}_{-0.49}$ (the dotted line in
Figure~\ref{fig:uqporin}; a systematic error of 2.3\% is added for
$R_{\rm MCDPS}$) using only data on the HB and
$\beta=-2.01^{+0.35}_{-0.51}$ (the dashed line; a systematic error of
2.1\% is added for $R_{\rm MCDPS}$) if data on the NB are also
used. These values are consistent with $-1.5$ (the solid line, forced
to pass the data point with $S_{\rm Z}$=0.96--1.00), expected if the
frequency of the upper kHz QPO is the Keplerian frequency at $R_{\rm
  MCDPS}$. The systematic error of $\beta$ due to our choice of
$\Gamma_{\rm nthComp}$ in Model~2 is estimated to be around 0.6 based
on comparison with results using other values of $\Gamma_{\rm
  nthComp}$ in the range of 2.0--2.7. In Model~1, $R_{\rm MCD}$ on the
HB increases with the frequency of the upper kHz QPO. Therefore, it is
important to note that the relationship between the upper kHz QPO
frequency and the inner disk radius changes sign between the use of
Models~1 and 2. The result shown in Figure~\ref{fig:uqporin} depends
on a model that counts the number of Comptonized photons, so that the
QPO frequency becomes correlated to the apparent inner radius of the
disk prior to the effects of Compton scattering.

Figure~\ref{fig:hborin} shows the frequency of the HBO versus $R_{\rm
  MCDPS}$ from Model~2. We see that the frequency of the HBO increases
as $R_{\rm MCDPS}$ decreases down the HB, but both become relatively
constant on the NB. We note that in fact the frequency of the HBO
decreases slightly on the lower NB
\citep[Figure~\ref{fig:fresz};][]{hovajo2002}. Also assuming a
relation of frequency~$\propto$~$R_{\rm MCDPS}^{\beta}$, we estimate
$\beta=-3.70^{+0.63}_{-0.96}$ (the dotted line in
Figure~\ref{fig:hborin}; a systematic error of 1.6\% added for $R_{\rm
  MCDPS}$) using only data on the HB and $\beta=-4.35^{+0.66}_{-0.94}$
(the dashed line; a systematic error of 2.8\% added for $R_{\rm
  MCDPS}$) if data on the NB are also used. These values are roughly
consistent with $-3$ (the solid line in Figure~\ref{fig:hborin}), a
value expected in a Lense-Thirring interpretation of the HBO
\citep{stvi1998,va2006}. Based on the results using other choices of
$\Gamma_{\rm nthComp}$, we estimate a systematic error of 1.2 for the
above measurements of $\beta$.

\subsection{The Source Luminosity Based on Type I X-ray Bursts}
\label{sec:discburst}
Type I X-ray bursts in NS LMXBs can provide critical reference
quantities such as the Eddington flux and the apparent NS surface area
when they show photospheric radius expansion. These quantities can be
compared with the spectral modeling results to infer the source
luminosity and the visible size of the boundary layers, which are key
elements in constructing the physical picture of the accretion process
\citep[][LRH09]{lireho2007,lireho2010}. \gx\ shows both short
($\sim$10 s) and long ($\gtrsim$100 s) bursts
\citep{kuhova2002,gamuha2008}. With the persistent emission comparable
to the peak net burst emission, these bursts are probably observed at
accretion rates around the Eddington limit and are not expected
\citep{gamuha2008,relico2006}. It is still under debate on how to
analyze bursts at such high accretion rates. In the ``standard'' burst
analysis, burst spectra are obtained by subtracting the persistent
emission from the total emission and then fitted with a simple BB
model. At high accretion rates, the persistent emission from the
boundary layer may be strong. If it varies during bursts, it is
instead probably more appropriate to fit the total emission with
multiple spectral components, typically including a BB to describe the
burst emission \citep[e.g.,][]{szvale1986}.

\citet{kuhova2002} applied the ``standard'' analysis to ten bursts of
\gx\ from \xte\ data and obtained successful fits for most of
them. Here we combine our spectral modeling results of the persistent
emission with their burst analysis to infer the Eddington flux and
apparent NS area. In our data set, there are no short bursts but five
long ones, corresponding to bursts 6--10 in \citet{kuhova2002}. Burst
10 is near the lower vertex ($S_{\rm Z}=2.18$) but shows no
photospheric radius expansion. Bursts 6--9 are near the upper vertex
($S_{\rm Z}=0.86$, 1.04, and 0.73, respectively) and show photospheric
radius expansion, and we focus on them. In these bursts, for several
tens of seconds right after the touch-down (the moment when the
expanded matter falls back onto the NS surface), the burst
temperatures peak and hover around $\simeq$2.65 keV, and the apparent
net burst area is roughly constant ($N_{\rm BB}$$\simeq$25),
corresponding to a peak net burst flux around 1.24$\times$10$^{-8}$
erg~cm$^{-2}$~s$^{-1}$ (from burst 6). In comparison, the boundary
layer during the persistent emission has a similar temperature and
$N_{\rm BB}$$\simeq$15 from our modeling, about a size of 60\% of the
net burst area.

With such a strong boundary layer, the Eddington flux and the apparent
NS area would be very different depending on whether we infer them
only from the net burst emission or have the boundary layer emission
also included. We note this is not a serious issue in atoll sources,
as their boundary layers are small ($\lesssim$10\% of the NS surface)
and have temperatures typically lower than the bursts
\citep[][LRH09]{lireho2007,lireho2010}. Considering that both the
boundary layer and the burst emission are from the NS surface, we add
them together and obtain the Eddington flux of about
2.0$\times$10$^{-8}$ erg~cm$^{-2}$~s$^{-1}$ and the total apparent NS
surface area of $N_{\rm BB}$$\simeq$40 for \gx, but we also discuss
the results with the Eddington flux and the total NS apparent area
calculated from the burst emission only.

The luminosities in Tables~\ref{tbl-fitparmod1}--\ref{tbl-fitparmod2}
are in units of the Eddington flux obtained above, and they would be
60\% larger if we do not include the boundary layer emission in the
calculation of the Eddington flux above. In the upper vertex, \gx\ has
luminosity of a factor of 1.5 of the Eddington limit, higher than
\object{XTE J1701-462} in the Sco-like stage (below 0.9 Eddington
luminosity (LRH09)). If $\dot{M}$ accounts for different source types
(Section~\ref{sec:interZ}), one might expect their luminosities to be
similar. The above difference might be due to several uncertain
factors, such as the inclination and the thickness of the accretion
flow not taken into account in the Eddington flux correction
above. However, we cannot rule out that their difference in luminosity
is real , which could be due to factors such as different compositions
of the accreted matterials and different NS masses.

We estimate the source distance based on the Eddington flux obtained
above. Using the empirical value of the Eddington luminosity of
3.79$\times$10$^{38}$ erg~s$^{-1}$ from \citet{kudein2003}, the source
distance is 12.6 kpc, which is assumed in this paper. If the
theoretical expression of the Eddington limit \citep[see Equation 8
  in][]{gamuha2008} is used and a NS with a mass of 1.4 \msun\ and a
radius of 10 km is assumed, we obtain a source distance of 10.5 kpc
for the H-poor case (the H-fraction $X=0$) and 8.1 kpc for the H-rich
case ($X=0.7$). The above distances should increase by 26\% if we only
use the net burst emission to estimate the Eddington flux.

\section{DISCUSSION}
\label{sec:dis}
\subsection{The Role of $\dot{M}$ in the Spectral Evolution}
\label{sec:interZ}
The spectral evolution of \gx\ along the Z track based on Model~1
(Figure~\ref{fig:Tlumm1}) is very similar to that of \object{XTE
  J1701-462} in the Sco-like stage as obtained by LRH09 using a
similar model. This supports the viability of physical interpretations
for Z branch evolution offered in that paper (see
Section~\ref{sec:reszbranch}). An associated question is whether the
$\dot{M}$ into the disk varies along the Z track. LRH09 inferred a
constant $\dot{M}$ along the Sco-like Z tracks of \object{XTE
  J1701-462} based on the MCD component for the NB and FB. Although in
the HB the MCD component did not follow a constant $\dot{M}$, the
variation of the MCD luminosity was largely offset by that of the
Comptonized component, leading LRH09 to surmise that on the HB
$\dot{M}$ is also constant, with the thermal disk emission simply
converted to the Comptonized emission as the source ascends the
HB. Similar behavior observed in \gx\ using Model~1 implies that
$\dot{M}$ is probably also constant in this system. Model~2 took a
step further by providing an empirically self-consistent way to model
the Comptonized component, assuming that a hot corona in the line of
sight to the disk Comptonizes part of the thermal disk emission. We
have seen an intriguing result of this model, i.e., $L_{\rm MCDPS}$,
the thermal disk luminosity prior to Compton scattering, is consistent
with lying along a constant $\dot{M}$ line over the whole Z track
(Figure~\ref{fig:Tlum}). We note that in the above we use the MCDPS to
infer the $\dot{M}$ into the disk. This assumes no significant mass
loss around the area where most of the thermal disk emission is
produced. At the very inner part of the disk, some of this $\dot{M}$
can be taken away by mass outflow (Section~\ref{sec:reszbranch}), with
the rest going onto the NS surface to form the boundary layer. Based
on the little variation of the total flux on the HB and NB seen in
\citet{distro2000} and the comparison with black hole X-ray binaries,
\citet{hovajo2002} also suggested a constant $\dot{M}$ along the Z
track.

The conclusion of a constant $\dot{M}$ into the disk along the Z track
is further supported by the global evolution of \gx. Its secular
changes were observed to be very small, either from our data set
spanning nine days or from data spanning years in \citet{wihova1997}
and \citet{hovajo2002}. This is in contrast with strong secular
changes seen in \object{XTE J1701-462} in 2006--2007, which were
ascribed to the variation in the $\dot{M}$ into the disk
\citep[LRH09,][]{hovawi2007,hovafr2010}. Then a reasonable explanation
for the lack of strong secular changes in \gx\ is that its $\dot{M}$
has been fairly constant. This interpretation implies a need of hardly
varying $\dot{M}$ over its entire Z track, in accordance with our
spectral fit results.

\subsection{The Interpretation of Z Branches}
\label{sec:reszbranch}

\begin{figure}
%\figurenum{9}  
\plotone{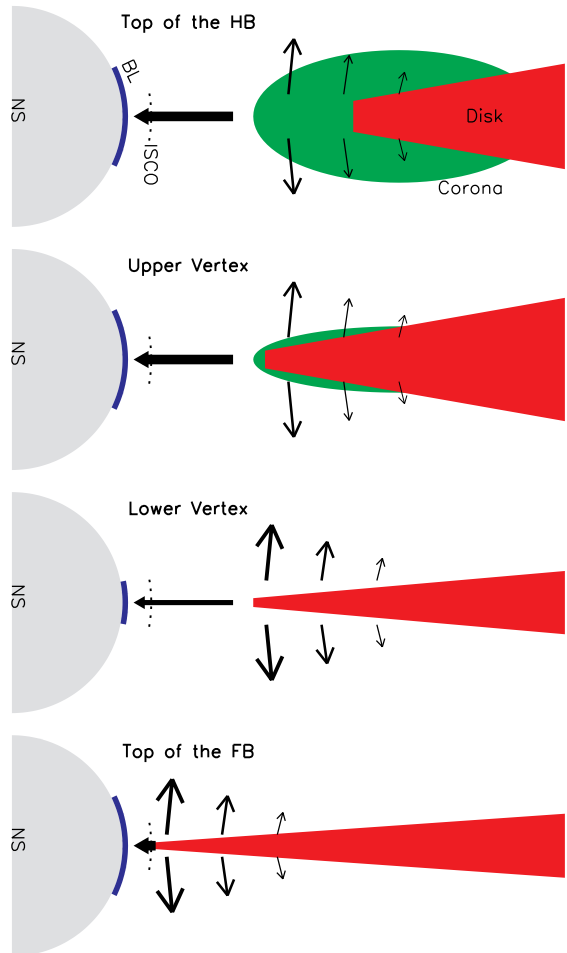} 
\caption{Schematic sketch of the accretion flow at different positions
  of the Z track. The system is assumed to be axisymmetric relative to
  the rotational axis, and the sketch shows one half of the
  cross-section through the rotational axis. Arrows represent the mass
  flow onto the NS surface and out of the disk (at the inner disk),
  with the strength roughly indicated by the thickness of the
  arrows. The blue arc represents the boundary layer (BL).
\label{fig:scheplot}} 
\end{figure}

If the $\dot{M}$ into the disk does not change along the Z track as
argued above, the remaining main question is what causes the evolution
along the Z track. Based on \object{XTE J1701-462}, LRH09 suggested
that the three Sco-like branches are due to three physical processes
operating at a constant $\dot{M}$ into the disk. The source climbs up
the HB due to the increasing portion of the thermal disk emission
Compton scattered. The excursion from the lower to the upper vertices
on the NB corresponds to the transition from a slim-disk accretion
\citep{abczla1988} to a standard-thin-disk accretion
\citep{shsy1973}. At the lower vertex, the disk is truncated at a
radius outside the ISCO as the local Eddington limit is reached at the
inner disk radius (see discussion below). The source climbs up the FB
due to the temporarily fast decrease of the inner disk radius toward
the ISCO. Since the source evolved slower and stayed longer in the two
vertices than in other parts of Z tracks (at least on the NB and FB),
the above three processes could correspond to three kinds of
instability (LRH09). The slim disk solution was associated with the
upper vertex because it can provide more mass inflow onto the NS
surface than the standard thin disk and can explain the increase of
the BB apparent area from the lower to the upper vertices. The disk
emission was not observed to change significantly during such a
transition (in the Sco-like tracks), probably because the slim disk
emission differs from the standard disk significantly only at
accretion rates much higher than the Eddington limit
\citep{mikata2000}.
 
As a similar spectral evolution is observed in \gx, the above picture
can be applied to this system and is sketched in
Figure~\ref{fig:scheplot}. The disk is drawn to be thicker for the HB
to represent a slim disk, and is thinner for the FB to represent a
standard thin disk. On the HB, the disk is surrounded by a
Comptonizing corona with a temperature of about 6 keV, based on our
fits with Model~2 in this work. Nearly one half of the thermal disk is
scattered at the top of the HB, but this quantity has a strong
dependence on $\Gamma_{\rm nthComp}$
(Section~\ref{sec:spectralfit}). The optical depth of the corona is
about a few from Model~2, based on its relation with $\Gamma_{\rm
  nthComp}$ and $kT_{\rm e,nthComp}$ \citep{zdjoma1996}. The
Comptonized component is strong only at the top of the HB, making it
hard to measure the variation of the optical depth along the Z
track. Thus we cannot easily infer whether the decrease of the
Comptonization emission as the source descends the HB is due to the
decrease of the covering fraction of the corona or due to the decrease
of its optical depth.

We add some small arrows in Figure~\ref{fig:scheplot} to indicate the
possible strong hot mass outflow at the very inner part of the
accretion disk in this system. This is expected to be a common
phenomenon for a disk accretion system at high accretion rates, in
which the mass can be input into the disk from outside and ejected as
a disk wind at the very inner part of the disk where the emission
perpendicular to the disk plan reaches the local Eddington limit
\citep{ka1980,wafuta2000,mikata2000,fu2004,ohmi2007}. The mass outflow
can also explain the low BB luminosity, relative to the disk
luminosity (discussed below).

Figure~\ref{fig:scheplot} does not include the jet component that is
possibly responsible for the radio emission detected from this
system. \citet{mimife2007} observed the radio emission to decrease
significantly from the HB to the FB along the Z track and found it to
correlate with a hard spectral component (the ``hard tail''), which
they fitted with a PL. In this paper we used a different dataset and a
different spectral model from those used by
\citet{mimife2007}. However, we found that consistent results can also
be obtained if we fitted our data with their model (the same as the
one used by \citet{distro2000}). The CPL/nthComp component in our
model was also observed to decrease significantly from the HB to the
FB, in a similar way as the hard tail in \citet{mimife2007}, thus
probably also correlating with the radio emission. It played a similar
role as the PL component in \citet{mimife2007} in that both of them
are the dominating spectral component at energies above 30 keV when
they are the strongest, at the top of the HB. We have shown that the
CPL/nthComp model described our data better than the PL model
(Figure~\ref{fig:comdelchi}). We note that the HB spectra in
\citet{mimife2007} seemed to extend above 60 keV, which was not seen
in our HB spectra. The 60--100 keV HEXTE count rates are about three
$\sigma$ and about 2\% of the background level in their HB observation
70023-01-01-00. To check whether the apparent lack of the hard tail
above 60 keV in our data is due to relatively small exposures of our
$S_{\rm Z}$-resolved spectra (each with a few ks at the top of the HB,
compared to $\sim$10 ks for the spectrum in \citet{mimife2007}), we
combined the spectra of the first five $S_{\rm Z}$ selections
(representing the upper half of the HB, where the hard tail was
reported to be strongest), resulting in a spectrum with an exposure
time of $\sim$20 ks. We found that the source was detected at less
than one $\sigma$ in the 60--100 keV band (combining data from both
HEXTE clusters), suggesting that this (additional) hard emission may
be transient, even for the same location in the Z track.

The hard tail observed by {\it BeppoSAX} from \gx\ in the HB has been
explained in two different models, i.e., hybrid Comptonization and
bulk motion Comptonization, by \citet{fafrza2005} and
\citet{fatifr2007}, respectively. In the hybrid Comptonization model
(using the {\it eqpair} model \citep{co1999} in XSPEC), two electron
populations (thermal and non-thermal) are required, with the
non-thermal one accounting for most of the hard tail above $\sim$30
keV. In the bulk motion Comptonization model (using the {\it bmc}
model \citep{timaky1997} in XSPEC), the hard tail is caused by Compton
upscattering of soft photons by energetic electrons in a converging
flow onto the NS. The hard component (CPL/nthComp) in our HB spectra
has an exponential cutoff at $\sim$10 keV and has been explained by us
as due to Comptonization by a low-temperature ($\sim$6 keV) thermal
plasma covering the accretion disk (i.e., the non-thermal electron
population is not needed). The bulk motion Comptonization model cannot
explain this component easily because the cutoff energy of its
emergent spectrum is much higher \citep[a few hundred
  keV,][]{timaky1997}.

In Figure~\ref{fig:scheplot} the boundary layer, which we model with a
BB, is pictured as an equatorial belt on the NS surface. On the HB,
$N_{\rm BB}$ of the boundary layer is around 15, while the entire NS
surface has a corresponding value of 40 if we sum the boundary layer
area and the burst emission area together
(Section~\ref{sec:discburst}), indicating that the boundary layer has
an apparent fraction of 3/8 of the entire NS surface. The real
fraction is different, because of the special geometry of the boundary
layer pictured above, and can be estimated following
\citet{lireho2007}. We infer the maximal latitude of the boundary
layer to be 26$\degr$ and 35$\degr$, corresponding to 43\% and 57\% of
the entire NS surface, for an inclination of 60$\degr$ and 30$\degr$,
respectively. In the lower vertex, the boundary layer has $N_{\rm
  BB}$$\sim$6, and we infer the maximal latitude of the boundary layer
to be 11$\degr$ and 18$\degr$, corresponding to 19\% and 30\% of the
entire NS surface, for an inclination of 60$\degr$ and 30$\degr$,
respectively. The solutions for an inclination of 60$\degr$ are used
to create Figure~\ref{fig:scheplot}, except for the panel for the top
of the FB, which is arbitrary due to large uncertainty. The above
estimates of the size of the boundary layer are only approximate,
neglecting factors such as the thickness of accretion flow, and should
be larger if we estimate the NS surface area only from the burst
emission area. The above corrections imply that the real boundary
layer luminosities should be higher than those shown in
Figures~\ref{fig:Tlum}--\ref{fig:parameters} by up to 50\%. However,
even with such corrections, the luminosity of the boundary layer seems
to be still less than the disk luminosity, which is not expected from
the simple energy argument \citep{miinko1984}. One explanation for
this is the mass outflow described above.

Although large uncertainties are seen on the FB
(Section~\ref{sec:spev}), the boundary layer on the HB and NB has
temperatures around 2.6 keV, about the peak value seen in the radius
expansion bursts from this system \citep{kuhova2002}. This indicates
that the boundary layer emission may have reached the local Eddington
limit. In such a situation, the increase in the mass accretion rate
onto the NS surface will just lead to increase in the boundary layer
emission area (not its temperature), as seen on the NB. This was also
observed in \object{XTE J1701-462} (LRH09).

\subsection{The Origins of kHz QPOs and HBOs}
We have found that the dependence of the frequency of the upper kHz
QPO on the apparent inner disk radius $R_{\rm MCDPS}$ is roughly
consistent with frequency~$\propto$~$R_{\rm MCDPS}^{-3/2}$ from
Model~2, making it natural to identify it as the Keplerian frequency
at the inner disk radius. This has been a critical assumption in many
explanations of the kHz QPOs \citep{liboba2011,va2006}. We note that
$R_{\rm MCDPS}$ has been corrected for the photons which we assumed to
be from the thermal disk emission but scattered by the corona. We also
caution that the absolute values of the radius measurements have scale
uncertainties that depend on the source distance, the inclination, the
hardening effect, and the NS mass \citep{zhcuch1997}. There is another
important factor \citep[$\eta$ in][]{zhcuch1997} to account for the
difference between the real inner disk radius and the radius where the
disk temperature peaks. All these factors have too large uncertainties
to allow for a meaningful estimate of the real inner disk radius
directly from $R_{\rm MCDPS}$. We calculated the variation of the
upper kHz QPO with $R_{\rm MCDPS}$ to infer its dependence on the real
inner disk radius, assuming that the above conversion factors bewteen
$R_{\rm MCDPS}$ and the real inner disk radius are fairly constant, at
least along the Z track where upper kHz QPOs are observed.

The frequency of the HBO is too low to be the Keplerian frequency at
the inner disk radius. However, we also observe its close dependence
on $R_{\rm MCDPS}$ covering the entire HB and NB. Thus the HBO is
probably intimately related to the dynamics at the inner disk radius
in some way. In the Lense-Thirring interpretation for the HBO, an
approximate relation of frequency~$\propto$~$R_{\rm MCDPS}^{-3}$ is
expected \citep{stvi1998,va2006}, and our results are roughly
consistent with this. Although the frequency of the HBO varies much
less on the NB than on the HB, it still shows a small increase first
and then a small decreases as the source descends the NB
\citep[Figure~\ref{fig:fresz};][]{hovajo2002}. The error bars of
$R_{\rm MCDPS}$ are not small enough for us to conclude whether it
correlates with this evolution pattern. However, we cannot rule out
that such an evolution of the HBO on the NB is due to the change of
the disk structure from a slim disk to a standard thin disk on the NB
(Section 5.3). The decrease in the quality of the HBO on the lower NB
\citep{hovajo2002} might be related to this.

\section{CONCLUSIONS}
We have analyzed a Z track of \gx\ from \xte\ observations between
1999 October 3--12. Our analysis has confirmed similar spectral
properties of \gx\ to those of \object{XTE J1701-462} in the Sco-like
stage. In this study, we took a further step by modeling the
Comptonized component in an empirically self-consistent way and found
that the evolution of the thermal disk is consistent with being at a
constant $\dot{M}$ into it over the whole track, supporting our
conclusion of a constant $\dot{M}$ over the whole Sco-like Z
track. Similar to \object{XTE J1701-462}, we found that the branches
of \gx\ can be explained as being due to three processes operating at a
constant $\dot{M}$ into the disk. The HB is due to increase of the
Comptonization with respect to the upper vertex. The FB is due to
shrinking the inner disk radius on fast timescales from the lower
vertex. on the NB we saw variation in the area of the boundary layer,
which can be explained as being due to transition from accretion through a
slim disk in the upper vertex to a standard thin disk in the lower
vertex. The boundary layer has a size of $\sim$20--30\% of the entire
NS surface in the lower vertex and $\sim$40--60\% in the upper vertex
and is probably at the local Eddington limit.

The dependence of the frequency of the upper kHz QPO on the apparent
inner disk radius is found to be roughly consistent with
frequency~$\propto$~$R_{\rm MCDPS}^{-3/2}$ on the HB, supporting
identification of the frequency of the upper kHz QPO as the Keplerian
frequency at the inner disk radius. We measure
frequency~$\propto$~$R_{\rm MCDPS}^{-4.0\pm0.7}$ for the HBO over the
entire HB and NB, indicating an intimate relation between the HBO and
the dynamics at the inner disk.

\acknowledgments 

Acknowledgments: DL thanks Piotr Zycki, James Steiner, and Joseph
Neilsen for helpful comments on Comptonization models.

\tabletypesize{\scriptsize}
\begin{deluxetable}{ccccccccccccc}
\setlength{\tabcolsep}{0.01in}
\tablecaption{Spectral fit results using Model~1 \label{tbl-fitparmod1}}
\tablewidth{0pt}
\tablehead{ \colhead{$S_{\rm Z}$} & \colhead{EXP} & \colhead{$kT_{\rm MCD}$} &\colhead{$N_{\rm MCD}$} & 
  \colhead{$kT_{\rm BB}$} &\colhead{$N_{\rm BB}$} &
  \colhead{$N_{\rm CPL}$} &
  \colhead{$\sigma_{\rm line}$} & \colhead{$N_{\rm line}$} &\colhead{EW}&\colhead{$\tau_{\rm edge}$}&
  \colhead{$\chi^2_\nu(\nu)$} & \colhead{$L_{\rm X, Edd}$}\\
           & (ks) & (keV) & &
  (keV) & &
  &
   (keV) &($10^{-2}$) & (eV) &($10^{-2}$) & &\\
(1)&(2)&(3)&(4)&(5)&(6)&(7)&(8)&(9)&(10)&(11)&(12)&(13)
}
\startdata
0.07--0.24 & 2.1   & $ 1.52$$\pm$$ 0.05$  & $ 85$$\pm$$  7$  & $ 2.66$$\pm$$ 0.04$  & $ 14.7$$\pm$$  0.9$  & $ 1.58$$\pm$$ 0.11$  & $ 0.36$$\pm$$ 0.19$  & $1.2$$\pm$$0.3$  & $ 77$$\pm$$ 18$  & $2.9$$\pm$$1.4$ &0.75(151)&$ 1.44\pm 0.06$\\
0.24--0.32 & 2.3   & $ 1.56$$\pm$$ 0.04$  & $ 91$$\pm$$  7$  & $ 2.69$$\pm$$ 0.04$  & $ 14.6$$\pm$$  0.9$  & $ 1.30$$\pm$$ 0.11$  & $ 0.42$$\pm$$ 0.20$  & $1.2^{+0.5}_{-0.3}$  & $ 76$$\pm$$ 20$  & $2.5$$\pm$$1.5$ &0.78(149)&$ 1.46\pm 0.05$\\
0.32--0.40 & 5.7   & $ 1.58$$\pm$$ 0.03$  & $ 95$$\pm$$  6$  & $ 2.66$$\pm$$ 0.03$  & $ 15.3$$\pm$$  0.8$  & $ 1.23$$\pm$$ 0.07$  & $ 0.40$$\pm$$ 0.19$  & $1.1$$\pm$$0.3$  & $ 66$$\pm$$ 17$  & $3.1$$\pm$$1.2$ &0.80(154)&$ 1.48\pm 0.04$\\
0.40--0.48 & 6.4   & $ 1.59$$\pm$$ 0.03$  & $ 97$$\pm$$  5$  & $ 2.66$$\pm$$ 0.03$  & $ 15.3$$\pm$$  0.8$  & $ 1.13$$\pm$$ 0.07$  & $ 0.36$$\pm$$ 0.18$  & $1.1$$\pm$$0.3$  & $ 63$$\pm$$ 16$  & $3.2$$\pm$$1.2$ &0.86(155)&$ 1.49\pm 0.04$\\
0.48--0.56 & 3.1   & $ 1.63$$\pm$$ 0.03$  & $ 98$$\pm$$  6$  & $ 2.70$$\pm$$ 0.04$  & $ 14.5$$\pm$$  0.8$  & $ 0.92$$\pm$$ 0.10$  & $ 0.43$$\pm$$ 0.21$  & $1.2^{+0.5}_{-0.3}$  & $ 72$$\pm$$ 18$  & $2.6$$\pm$$1.5$ &0.96(146)&$ 1.49\pm 0.05$\\
0.56--0.64 & 3.0   & $ 1.64$$\pm$$ 0.03$  & $106$$\pm$$  5$  & $ 2.68$$\pm$$ 0.04$  & $ 15.2$$\pm$$  0.9$  & $ 0.71$$\pm$$ 0.09$  & $ 0.33$$\pm$$ 0.21$  & $1.0$$\pm$$0.3$  & $ 56$$\pm$$ 14$  & $2.5$$\pm$$1.2$ &0.98(147)&$ 1.51\pm 0.04$\\
0.64--0.72 & 8.5   & $ 1.64$$\pm$$ 0.02$  & $108$$\pm$$  4$  & $ 2.65$$\pm$$ 0.03$  & $ 15.4$$\pm$$  0.8$  & $ 0.71$$\pm$$ 0.06$  & $ 0.36$$\pm$$ 0.16$  & $1.0$$\pm$$0.2$  & $ 60$$\pm$$ 13$  & $3.0$$\pm$$1.1$ &1.01(147)&$ 1.52\pm 0.03$\\
0.72--0.80 & 16.7   & $ 1.64$$\pm$$ 0.02$  & $111$$\pm$$  4$  & $ 2.60$$\pm$$ 0.02$  & $ 16.3$$\pm$$  0.8$  & $ 0.74$$\pm$$ 0.05$  & $ 0.30$$\pm$$ 0.18$  & $1.0$$\pm$$0.2$  & $ 55$$\pm$$ 14$  & $3.8$$\pm$$1.0$ &1.13(139)&$ 1.53\pm 0.03$\\
0.80--0.88 & 20.0   & $ 1.66$$\pm$$ 0.02$  & $113$$\pm$$  4$  & $ 2.60$$\pm$$ 0.02$  & $ 16.3$$\pm$$  0.8$  & $ 0.63$$\pm$$ 0.05$  & $ 0.31$$\pm$$ 0.17$  & $1.0$$\pm$$0.2$  & $ 57$$\pm$$ 13$  & $3.7$$\pm$$1.0$ &1.06(142)&$ 1.55\pm 0.03$\\
0.88--0.96 & 24.8   & $ 1.67$$\pm$$ 0.02$  & $113$$\pm$$  4$  & $ 2.59$$\pm$$ 0.02$  & $ 16.3$$\pm$$  0.8$  & $ 0.56$$\pm$$ 0.04$  & $ 0.27^{+ 0.15}_{-0.26}$  & $1.0$$\pm$$0.2$  & $ 54$$\pm$$ 12$  & $4.3$$\pm$$1.0$ &1.03(140)&$ 1.56\pm 0.02$\\
0.96--1.00 & 9.1   & $ 1.71$$\pm$$ 0.02$  & $112$$\pm$$  4$  & $ 2.62$$\pm$$ 0.03$  & $ 14.9$$\pm$$  0.9$  & $ 0.40$$\pm$$ 0.06$  & $ 0.30^{+ 0.15}_{-0.22}$  & $1.0$$\pm$$0.2$  & $ 54$$\pm$$ 15$  & $3.7$$\pm$$1.0$ &0.88(146)&$ 1.55\pm 0.03$\\
1.00--1.04 & 8.0   & $ 1.70$$\pm$$ 0.02$  & $112$$\pm$$  4$  & $ 2.60$$\pm$$ 0.03$  & $ 15.1$$\pm$$  1.0$  & $ 0.42$$\pm$$ 0.06$  & $ 0.31$$\pm$$ 0.18$  & $1.0$$\pm$$0.2$  & $ 55$$\pm$$ 14$  & $3.8$$\pm$$1.1$ &0.87(147)&$ 1.55\pm 0.03$\\
1.04--1.12 & 15.7   & $ 1.70$$\pm$$ 0.02$  & $117$$\pm$$  4$  & $ 2.57$$\pm$$ 0.03$  & $ 15.5$$\pm$$  0.9$  & $ 0.37$$\pm$$ 0.04$  & $ 0.27^{+ 0.15}_{-0.26}$  & $0.9$$\pm$$0.2$  & $ 52$$\pm$$ 13$  & $4.2$$\pm$$1.0$ &1.10(143)&$ 1.54\pm 0.02$\\
1.12--1.20 & 20.8   & $ 1.71$$\pm$$ 0.02$  & $117$$\pm$$  3$  & $ 2.59$$\pm$$ 0.03$  & $ 14.2$$\pm$$  0.8$  & $ 0.25$$\pm$$ 0.04$  & $ 0.25^{+ 0.15}_{-0.25}$  & $0.9$$\pm$$0.2$  & $ 53$$\pm$$ 12$  & $4.0$$\pm$$1.0$ &1.04(141)&$ 1.53\pm 0.02$\\
1.20--1.28 & 16.0   & $ 1.71$$\pm$$ 0.02$  & $118$$\pm$$  4$  & $ 2.57$$\pm$$ 0.03$  & $ 13.6$$\pm$$  0.9$  & $ 0.25$$\pm$$ 0.04$  & $ 0.27^{+ 0.15}_{-0.26}$  & $0.9$$\pm$$0.2$  & $ 52$$\pm$$ 12$  & $4.4$$\pm$$1.0$ &0.98(144)&$ 1.51\pm 0.02$\\
1.28--1.36 & 6.9   & $ 1.71$$\pm$$ 0.02$  & $118$$\pm$$  4$  & $ 2.58$$\pm$$ 0.04$  & $ 12.8$$\pm$$  1.0$  & $ 0.20$$\pm$$ 0.06$  & $ 0.29^{+ 0.15}_{-0.24}$  & $0.9$$\pm$$0.2$  & $ 52$$\pm$$ 13$  & $4.5$$\pm$$1.1$ &0.94(134)&$ 1.48\pm 0.03$\\
1.36--1.44 & 6.5   & $ 1.70$$\pm$$ 0.02$  & $118$$\pm$$  4$  & $ 2.52$$\pm$$ 0.04$  & $ 12.9$$\pm$$  1.2$  & $ 0.26$$\pm$$ 0.06$  & $ 0.28^{+ 0.14}_{-0.21}$  & $0.9$$\pm$$0.2$  & $ 57$$\pm$$ 14$  & $5.7$$\pm$$1.1$ &0.99(136)&$ 1.44\pm 0.03$\\
1.44--1.52 & 6.1   & $ 1.68$$\pm$$ 0.02$  & $120$$\pm$$  4$  & $ 2.49$$\pm$$ 0.05$  & $ 12.6$$\pm$$  1.3$  & $ 0.29$$\pm$$ 0.05$  & $ 0.28$$\pm$$ 0.16$  & $1.0$$\pm$$0.2$  & $ 62$$\pm$$ 14$  & $5.5$$\pm$$1.1$ &1.12(141)&$ 1.41\pm 0.03$\\
1.52--1.60 & 5.2   & $ 1.69$$\pm$$ 0.02$  & $118$$\pm$$  4$  & $ 2.55$$\pm$$ 0.05$  & $ 10.7$$\pm$$  1.1$  & $ 0.20$$\pm$$ 0.06$  & $ 0.30$$\pm$$ 0.14$  & $1.1$$\pm$$0.2$  & $ 70$$\pm$$ 14$  & $5.1$$\pm$$1.2$ &1.02(134)&$ 1.37\pm 0.03$\\
1.60--1.68 & 5.8   & $ 1.70$$\pm$$ 0.02$  & $115$$\pm$$  4$  & $ 2.57$$\pm$$ 0.05$  & $  9.3$$\pm$$  1.0$  & $ 0.20$$\pm$$ 0.06$  & $ 0.31$$\pm$$ 0.14$  & $1.1$$\pm$$0.2$  & $ 73$$\pm$$ 14$  & $5.0$$\pm$$1.2$ &1.22(131)&$ 1.34\pm 0.03$\\
1.68--1.76 & 3.0   & $ 1.67$$\pm$$ 0.02$  & $121$$\pm$$  4$  & $ 2.54$$\pm$$ 0.06$  & $  9.7$$\pm$$  1.2$  & $ 0.15$$\pm$$ 0.07$  & $ 0.34$$\pm$$ 0.11$  & $1.3$$\pm$$0.2$  & $ 86$$\pm$$ 16$  & $5.4$$\pm$$1.3$ &1.26(126)&$ 1.31\pm 0.03$\\
1.76--1.84 & 3.1   & $ 1.68$$\pm$$ 0.02$  & $116$$\pm$$  4$  & $ 2.56$$\pm$$ 0.07$  & $  8.4$$\pm$$  1.1$  & $ 0.15$$\pm$$ 0.07$  & $ 0.35$$\pm$$ 0.11$  & $1.3$$\pm$$0.2$  & $ 95$$\pm$$ 16$  & $5.7$$\pm$$1.3$ &0.99(130)&$ 1.27\pm 0.03$\\
1.84--1.92 & 2.6   & $ 1.70$$\pm$$ 0.02$  & $112$$\pm$$  4$  & $ 2.64$$\pm$$ 0.07$  & $  6.9^{+  1.0}_{ -0.5}$  & $ 0.05^{+ 0.08}_{-0.05}$  & $ 0.35$$\pm$$ 0.10$  & $1.4$$\pm$$0.2$  & $105$$\pm$$ 17$  & $5.6$$\pm$$1.4$ &0.98(120)&$ 1.23\pm 0.03$\\
1.92--2.00 & 0.7   & $ 1.72^{+ 0.02}_{-0.04}$  & $107^{+  8}_{ -5}$  & $ 2.67^{+ 0.08}_{-0.14}$  & $  5.8^{+  1.6}_{ -0.9}$  & $ 0.06^{+ 0.14}_{-0.06}$  & $ 0.35$$\pm$$ 0.13$  & $1.5$$\pm$$0.3$  & $113$$\pm$$ 22$  & $6.3$$\pm$$2.1$ &0.90(111)&$ 1.21\pm 0.05$\\
2.00--2.08 & 2.3   & $ 1.70$$\pm$$ 0.02$  & $112$$\pm$$  4$  & $ 2.61$$\pm$$ 0.07$  & $  6.9$$\pm$$  0.9$  & $ 0.04^{+ 0.08}_{-0.04}$  & $ 0.35$$\pm$$ 0.09$  & $1.6$$\pm$$0.2$  & $118$$\pm$$ 17$  & $6.1$$\pm$$1.5$ &1.02(122)&$ 1.21\pm 0.03$\\
2.08--2.16 & 7.6   & $ 1.71$$\pm$$ 0.02$  & $111$$\pm$$  3$  & $ 2.62^{+ 0.03}_{-0.05}$  & $  7.1$$\pm$$  0.7$  & $ 0.04$$\pm$$ 0.04$  & $ 0.34$$\pm$$ 0.08$  & $1.6$$\pm$$0.2$  & $118$$\pm$$ 14$  & $5.8$$\pm$$1.2$ &1.39(128)&$ 1.23\pm 0.02$\\
2.16--2.24 & 6.5   & $ 1.73$$\pm$$ 0.02$  & $107$$\pm$$  3$  & $ 2.55$$\pm$$ 0.06$  & $  8.2$$\pm$$  1.0$  & $ 0.13$$\pm$$ 0.05$  & $ 0.33$$\pm$$ 0.08$  & $1.7$$\pm$$0.2$  & $117$$\pm$$ 14$  & $7.1$$\pm$$1.2$ &1.17(131)&$ 1.27\pm 0.03$\\
2.24--2.32 & 5.4   & $ 1.75$$\pm$$ 0.02$  & $103$$\pm$$  4$  & $ 2.58$$\pm$$ 0.06$  & $  8.3$$\pm$$  1.0$  & $ 0.08$$\pm$$ 0.06$  & $ 0.40$$\pm$$ 0.08$  & $1.9$$\pm$$0.2$  & $128$$\pm$$ 17$  & $6.2$$\pm$$1.3$ &1.27(127)&$ 1.30\pm 0.03$\\
2.32--2.40 & 3.8   & $ 1.82$$\pm$$ 0.02$  & $ 93$$\pm$$  3$  & $ 2.64$$\pm$$ 0.08$  & $  7.4$$\pm$$  1.1$  & $ 0.08$$\pm$$ 0.08$  & $ 0.34$$\pm$$ 0.08$  & $1.9$$\pm$$0.2$  & $123$$\pm$$ 14$  & $7.0$$\pm$$1.2$ &1.15(125)&$ 1.34\pm 0.04$\\
2.40--2.48 & 3.0   & $ 1.84$$\pm$$ 0.03$  & $ 93$$\pm$$  3$  & $ 2.66^{+ 0.04}_{-0.08}$  & $  7.7^{+  1.3}_{ -0.6}$  & $ 0.02^{+ 0.09}_{-0.02}$  & $ 0.39$$\pm$$ 0.08$  & $2.1$$\pm$$0.3$  & $128$$\pm$$ 18$  & $6.2$$\pm$$1.3$ &1.33(123)&$ 1.38\pm 0.03$\\
2.48--2.56 & 3.2   & $ 1.88$$\pm$$ 0.03$  & $ 87$$\pm$$  3$  & $ 2.67$$\pm$$ 0.08$  & $  7.5$$\pm$$  1.2$  & $ 0.08$$\pm$$ 0.08$  & $ 0.35$$\pm$$ 0.08$  & $2.0$$\pm$$0.2$  & $119$$\pm$$ 16$  & $7.0$$\pm$$1.2$ &1.44(130)&$ 1.41\pm 0.04$\\
2.56--2.64 & 1.9   & $ 1.92$$\pm$$ 0.03$  & $ 83^{+  3}_{ -1}$  & $ 2.76^{+ 0.05}_{-0.08}$  & $  6.7^{+  1.2}_{ -0.8}$  & $ 0.00^{+ 0.10}$  & $ 0.36$$\pm$$ 0.08$  & $2.1$$\pm$$0.2$  & $118$$\pm$$ 16$  & $7.6$$\pm$$1.3$ &1.31(129)&$ 1.44\pm 0.03$\\
2.64--2.72 & 2.6   & $ 1.96$$\pm$$ 0.03$  & $ 78$$\pm$$  3$  & $ 2.75^{+ 0.05}_{-0.10}$  & $  6.6^{+  1.4}_{ -0.9}$  & $ 0.05^{+ 0.11}_{-0.05}$  & $ 0.35$$\pm$$ 0.09$  & $2.0$$\pm$$0.3$  & $111$$\pm$$ 13$  & $7.0$$\pm$$1.3$ &1.44(128)&$ 1.48\pm 0.04$\\
2.72--2.80 & 3.1   & $ 1.99$$\pm$$ 0.02$  & $ 76$$\pm$$  3$  & $ 2.80^{+ 0.04}_{-0.07}$  & $  6.4$$\pm$$  0.8$  & $ 0.00^{+ 0.07}$  & $ 0.34$$\pm$$ 0.09$  & $2.1$$\pm$$0.3$  & $112$$\pm$$ 14$  & $6.8$$\pm$$1.1$ &1.41(128)&$ 1.51\pm 0.03$\\
2.80--2.88 & 2.5   & $ 2.00$$\pm$$ 0.03$  & $ 75$$\pm$$  3$  & $ 2.77$$\pm$$ 0.05$  & $  7.2$$\pm$$  1.0$  & $ 0.00^{+ 0.06}$  & $ 0.36^{+ 0.08}_{-0.05}$  & $2.2$$\pm$$0.3$  & $110$$\pm$$ 16$  & $6.7$$\pm$$1.2$ &1.41(128)&$ 1.55\pm 0.03$\\
2.88--2.96 & 2.3   & $ 2.04$$\pm$$ 0.03$  & $ 71^{+  3}_{ -1}$  & $ 2.81^{+ 0.05}_{-0.09}$  & $  6.7^{+  1.4}_{ -0.9}$  & $ 0.00^{+ 0.10}$  & $ 0.36$$\pm$$ 0.09$  & $2.2$$\pm$$0.3$  & $111$$\pm$$ 15$  & $6.8$$\pm$$1.3$ &1.25(130)&$ 1.58\pm 0.04$\\
2.96--3.04 & 1.9   & $ 2.06$$\pm$$ 0.03$  & $ 69$$\pm$$  3$  & $ 2.81$$\pm$$ 0.05$  & $  7.0$$\pm$$  1.1$  & $ 0.00^{+ 0.04}$  & $ 0.32$$\pm$$ 0.10$  & $2.2$$\pm$$0.3$  & $105$$\pm$$ 15$  & $7.2$$\pm$$1.2$ &1.42(122)&$ 1.62\pm 0.03$\\
3.04--3.12 & 1.9   & $ 2.10$$\pm$$ 0.04$  & $ 66^{+  3}_{ -2}$  & $ 2.82$$\pm$$ 0.07$  & $  6.8$$\pm$$  1.3$  & $ 0.00^{+ 0.10}$  & $ 0.38$$\pm$$ 0.10$  & $2.2$$\pm$$0.3$  & $104$$\pm$$ 17$  & $7.1$$\pm$$1.2$ &1.33(123)&$ 1.65\pm 0.04$\\
3.12--3.20 & 1.4   & $ 2.16$$\pm$$ 0.04$  & $ 62$$\pm$$  3$  & $ 2.89^{+ 0.08}_{-0.04}$  & $  5.8^{+  0.6}_{ -1.1}$  & $ 0.00^{+ 0.05}$  & $ 0.35$$\pm$$ 0.09$  & $2.2$$\pm$$0.3$  & $ 97$$\pm$$ 15$  & $7.3$$\pm$$1.3$ &1.25(125)&$ 1.69\pm 0.03$\\
3.20--3.40 & 2.5   & $ 2.17$$\pm$$ 0.04$  & $ 62$$\pm$$  3$  & $ 2.86$$\pm$$ 0.06$  & $  6.8$$\pm$$  1.2$  & $ 0.00^{+ 0.05}$  & $ 0.38$$\pm$$ 0.10$  & $2.3$$\pm$$0.4$  & $100$$\pm$$ 18$  & $7.0$$\pm$$1.3$ &1.39(126)&$ 1.77\pm 0.04$\\
3.40--3.60 & 2.0   & $ 2.27$$\pm$$ 0.05$  & $ 55^{+  4}_{ -2}$  & $ 2.98$$\pm$$ 0.08$  & $  5.3$$\pm$$  1.3$  & $ 0.00^{+ 0.07}$  & $ 0.38$$\pm$$ 0.10$  & $2.3$$\pm$$0.4$  & $ 94$$\pm$$ 15$  & $6.4$$\pm$$1.2$ &1.26(131)&$ 1.85\pm 0.05$\\
3.60--3.80 & 1.3   & $ 2.36$$\pm$$ 0.06$  & $ 51$$\pm$$  3$  & $ 3.04$$\pm$$ 0.12$  & $  4.4$$\pm$$  1.5$  & $ 0.00^{+ 0.06}$  & $ 0.39$$\pm$$ 0.10$  & $2.2$$\pm$$0.4$  & $ 84$$\pm$$ 18$  & $6.3$$\pm$$1.3$ &1.33(118)&$ 1.95\pm 0.06$\\
3.80--4.00 & 1.1   & $ 2.44$$\pm$$ 0.07$  & $ 47^{+  4}_{ -2}$  & $ 3.09$$\pm$$ 0.17$  & $  4.2^{+  2.0}_{ -1.1}$  & $ 0.00^{+ 0.10}$  & $ 0.39$$\pm$$ 0.10$  & $2.4$$\pm$$0.4$  & $ 87$$\pm$$ 17$  & $6.5$$\pm$$1.2$ &1.25(128)&$ 2.06\pm 0.08$\\
4.00--4.20 & 0.9   & $ 2.55$$\pm$$ 0.08$  & $ 42$$\pm$$  3$  & $ 3.22$$\pm$$ 0.21$  & $  3.2^{+  1.9}_{ -1.1}$  & $ 0.00^{+ 0.07}$  & $ 0.41$$\pm$$ 0.11$  & $2.4$$\pm$$0.5$  & $ 83$$\pm$$ 18$  & $5.2$$\pm$$1.3$ &1.38(121)&$ 2.14\pm 0.08$\\
4.20--4.40 & 1.0   & $ 2.68^{+ 0.06}_{-0.13}$  & $ 37^{+  5}_{ -2}$  & $ 3.34$$\pm$$ 0.27$  & $  2.3^{+  3.0}_{ -1.2}$  & $ 0.00^{+ 0.08}$  & $ 0.44$$\pm$$ 0.11$  & $2.5$$\pm$$0.5$  & $ 82$$\pm$$ 21$  & $5.3$$\pm$$1.1$ &1.12(125)&$ 2.23\pm 0.11$\\
4.40--4.80 & 1.1   & $ 2.90^{+ 0.06}_{-0.17}$  & $ 29^{+  5}_{ -1}$  & $ 3.53^{+ 0.78}_{-0.25}$  & $  1.0^{+  3.3}_{ -0.9}$  & $ 0.00^{+ 0.06}$  & $ 0.44$$\pm$$ 0.12$  & $2.5$$\pm$$0.3$  & $ 77$$\pm$$ 19$  & $4.2$$\pm$$1.1$ &1.02(127)&$ 2.34\pm 0.14$
\enddata 
\tablecomments{The first column is the $S_{\rm Z}$ range of each spectrum. The second column is the exposure per PCU. The last column is the total luminosity in units of the Eddington luminosity (3.79$\times$10$^{38}$ erg~s$^{-1}$; Section~\ref{sec:discburst}). See text for the meanings of other columns.}
\end{deluxetable}
%%%%\clearpage

\tabletypesize{\scriptsize}
\begin{deluxetable}{ccccccccccccc}
\setlength{\tabcolsep}{0.01in}
\tablecaption{Spectral fit results using Model~2\label{tbl-fitparmod2}}
\tablewidth{0pt}
\tablehead{ \colhead{$S_{\rm Z}$} & \colhead{$kT_{\rm MCD}$} &\colhead{$N_{\rm MCD}$} &\colhead{$N_{\rm MCDPS}$}& 
  \colhead{$kT_{\rm BB}$} &\colhead{$N_{\rm BB}$} &
  \colhead{$N_{\rm nthComp}$} &
  \colhead{$\sigma_{\rm line}$} & \colhead{$N_{\rm line}$} &\colhead{EW}&\colhead{$\tau_{\rm edge}$}&
  \colhead{$\chi^2_\nu(\nu)$} & \colhead{$L_{\rm X, Edd}$}\\
           & (keV) & & &
  (keV) & &
  &
   (keV) &($10^{-2}$) & (eV) &($10^{-2}$) & &\\
(1)&(2)&(3)&(4)&(5)&(6)&(7)&(8)&(9)&(10)&(11)&(12)&(13)
}
\startdata
0.07--0.24  & $ 1.32$$\pm$$ 0.04$  & $124$$\pm$$  9$  & $232$$\pm$$ 27$  & $ 2.59$$\pm$$ 0.04$  & $ 16.3^{+  0.9}_{ -1.4}$  & $ 0.94$$\pm$$ 0.09$  & $ 0.43^{+ 0.28}_{-0.17}$  & $1.4^{+1.1}_{-0.3}$  & $ 89$$\pm$$ 19$  & $3.3$$\pm$$1.5$ &0.83(151)&$ 1.48\pm 0.06$\\
0.24--0.32  & $ 1.42$$\pm$$ 0.04$  & $132^{+ 10}_{-15}$  & $195$$\pm$$ 19$  & $ 2.65^{+ 0.03}_{-0.06}$  & $ 15.4^{+  1.4}_{ -0.5}$  & $ 0.67$$\pm$$ 0.07$  & $ 0.81^{+ 0.14}_{-0.46}$  & $2.4^{+0.5}_{-1.2}$  & $148$$\pm$$ 43$  & $0.1^{+3.6}_{-0.1}$ &0.86(149)&$ 1.47\pm 0.05$\\
0.32--0.40  & $ 1.45$$\pm$$ 0.03$  & $124^{+ 11}_{ -7}$  & $178$$\pm$$ 14$  & $ 2.61^{+ 0.02}_{-0.03}$  & $ 16.7$$\pm$$  0.8$  & $ 0.62$$\pm$$ 0.05$  & $ 0.49^{+ 0.38}_{-0.19}$  & $1.3^{+1.1}_{-0.3}$  & $ 78$$\pm$$ 20$  & $2.9^{+1.4}_{-2.9}$ &0.86(154)&$ 1.50\pm 0.04$\\
0.40--0.48  & $ 1.49$$\pm$$ 0.03$  & $122$$\pm$$  6$  & $167$$\pm$$ 12$  & $ 2.61$$\pm$$ 0.03$  & $ 16.7$$\pm$$  0.8$  & $ 0.55$$\pm$$ 0.04$  & $ 0.42$$\pm$$ 0.20$  & $1.2^{+0.5}_{-0.3}$  & $ 71$$\pm$$ 18$  & $3.1$$\pm$$1.4$ &0.90(155)&$ 1.51\pm 0.04$\\
0.48--0.56  & $ 1.50^{+ 0.07}_{-0.03}$  & $131^{+  7}_{-17}$  & $165^{+ 12}_{-23}$  & $ 2.67$$\pm$$ 0.04$  & $ 15.5^{+  1.1}_{ -0.7}$  & $ 0.42$$\pm$$ 0.05$  & $ 0.81^{+ 0.14}_{-0.46}$  & $2.3^{+0.5}_{-1.2}$  & $138$$\pm$$ 43$  & $0.1^{+3.3}_{-0.1}$ &1.01(146)&$ 1.51\pm 0.05$\\
0.56--0.64  & $ 1.58$$\pm$$ 0.03$  & $122$$\pm$$  6$  & $144$$\pm$$ 10$  & $ 2.63$$\pm$$ 0.04$  & $ 16.3$$\pm$$  1.0$  & $ 0.32$$\pm$$ 0.05$  & $ 0.36$$\pm$$ 0.20$  & $1.0$$\pm$$0.3$  & $ 60$$\pm$$ 15$  & $2.4$$\pm$$1.3$ &0.99(147)&$ 1.52\pm 0.05$\\
0.64--0.72  & $ 1.58$$\pm$$ 0.03$  & $124$$\pm$$  5$  & $145$$\pm$$  8$  & $ 2.61$$\pm$$ 0.03$  & $ 16.5$$\pm$$  0.8$  & $ 0.31$$\pm$$ 0.03$  & $ 0.39$$\pm$$ 0.16$  & $1.1$$\pm$$0.3$  & $ 63$$\pm$$ 16$  & $2.8$$\pm$$1.1$ &0.97(147)&$ 1.53\pm 0.03$\\
0.72--0.80  & $ 1.58$$\pm$$ 0.02$  & $127$$\pm$$  5$  & $149$$\pm$$  7$  & $ 2.57$$\pm$$ 0.02$  & $ 17.5$$\pm$$  0.8$  & $ 0.32$$\pm$$ 0.02$  & $ 0.32$$\pm$$ 0.17$  & $1.0$$\pm$$0.2$  & $ 57$$\pm$$ 14$  & $3.7$$\pm$$1.0$ &1.06(139)&$ 1.54\pm 0.02$\\
0.80--0.88  & $ 1.61$$\pm$$ 0.02$  & $126$$\pm$$  5$  & $144$$\pm$$  6$  & $ 2.57$$\pm$$ 0.02$  & $ 17.3$$\pm$$  0.8$  & $ 0.26$$\pm$$ 0.02$  & $ 0.33$$\pm$$ 0.17$  & $1.1$$\pm$$0.2$  & $ 59$$\pm$$ 14$  & $3.5$$\pm$$1.0$ &1.01(142)&$ 1.55\pm 0.02$\\
0.88--0.96  & $ 1.63$$\pm$$ 0.02$  & $125$$\pm$$  4$  & $139$$\pm$$  6$  & $ 2.56$$\pm$$ 0.02$  & $ 17.3$$\pm$$  0.9$  & $ 0.23$$\pm$$ 0.02$  & $ 0.29^{+ 0.15}_{-0.24}$  & $1.0$$\pm$$0.2$  & $ 55$$\pm$$ 12$  & $4.1$$\pm$$1.0$ &0.98(140)&$ 1.56\pm 0.02$\\
0.96--1.00  & $ 1.67$$\pm$$ 0.02$  & $120$$\pm$$  5$  & $129$$\pm$$  6$  & $ 2.60$$\pm$$ 0.03$  & $ 15.7$$\pm$$  1.0$  & $ 0.16$$\pm$$ 0.02$  & $ 0.31$$\pm$$ 0.18$  & $1.0$$\pm$$0.2$  & $ 56$$\pm$$ 14$  & $3.6$$\pm$$1.0$ &0.84(146)&$ 1.56\pm 0.03$\\
1.00--1.04  & $ 1.67$$\pm$$ 0.02$  & $122$$\pm$$  5$  & $132$$\pm$$  6$  & $ 2.57$$\pm$$ 0.03$  & $ 16.1$$\pm$$  1.1$  & $ 0.17$$\pm$$ 0.02$  & $ 0.33$$\pm$$ 0.18$  & $1.0$$\pm$$0.2$  & $ 56$$\pm$$ 14$  & $3.7$$\pm$$1.1$ &0.83(147)&$ 1.55\pm 0.03$\\
1.04--1.12  & $ 1.67$$\pm$$ 0.02$  & $125$$\pm$$  4$  & $134$$\pm$$  5$  & $ 2.54$$\pm$$ 0.03$  & $ 16.4$$\pm$$  1.0$  & $ 0.15$$\pm$$ 0.02$  & $ 0.28^{+ 0.15}_{-0.28}$  & $1.0$$\pm$$0.2$  & $ 53$$\pm$$ 12$  & $4.1$$\pm$$1.0$ &1.03(143)&$ 1.55\pm 0.02$\\
1.12--1.20  & $ 1.69$$\pm$$ 0.02$  & $123$$\pm$$  4$  & $129$$\pm$$  5$  & $ 2.57$$\pm$$ 0.03$  & $ 14.9$$\pm$$  0.9$  & $ 0.10$$\pm$$ 0.02$  & $ 0.26^{+ 0.15}_{-0.26}$  & $1.0$$\pm$$0.2$  & $ 53$$\pm$$ 12$  & $4.0$$\pm$$1.0$ &0.98(141)&$ 1.53\pm 0.02$\\
1.20--1.28  & $ 1.69$$\pm$$ 0.02$  & $123$$\pm$$  4$  & $129$$\pm$$  5$  & $ 2.55$$\pm$$ 0.03$  & $ 14.3$$\pm$$  1.0$  & $ 0.10$$\pm$$ 0.02$  & $ 0.28^{+ 0.15}_{-0.28}$  & $0.9$$\pm$$0.2$  & $ 52$$\pm$$ 12$  & $4.3$$\pm$$1.0$ &0.93(144)&$ 1.51\pm 0.02$\\
1.28--1.36  & $ 1.69$$\pm$$ 0.02$  & $123$$\pm$$  5$  & $127$$\pm$$  6$  & $ 2.55$$\pm$$ 0.04$  & $ 13.5$$\pm$$  1.2$  & $ 0.08$$\pm$$ 0.02$  & $ 0.30^{+ 0.15}_{-0.23}$  & $0.9$$\pm$$0.2$  & $ 53$$\pm$$ 13$  & $4.5$$\pm$$1.1$ &0.90(134)&$ 1.48\pm 0.03$\\
1.36--1.44  & $ 1.67$$\pm$$ 0.03$  & $125$$\pm$$  5$  & $131$$\pm$$  6$  & $ 2.49$$\pm$$ 0.05$  & $ 13.7$$\pm$$  1.4$  & $ 0.11$$\pm$$ 0.02$  & $ 0.28^{+ 0.14}_{-0.21}$  & $1.0$$\pm$$0.2$  & $ 57$$\pm$$ 13$  & $5.7$$\pm$$1.1$ &0.94(136)&$ 1.44\pm 0.03$\\
1.44--1.52  & $ 1.66$$\pm$$ 0.03$  & $127$$\pm$$  5$  & $134$$\pm$$  6$  & $ 2.46$$\pm$$ 0.05$  & $ 13.6$$\pm$$  1.4$  & $ 0.12$$\pm$$ 0.02$  & $ 0.29^{+ 0.13}_{-0.19}$  & $1.0$$\pm$$0.2$  & $ 63$$\pm$$ 13$  & $5.5$$\pm$$1.1$ &1.06(141)&$ 1.42\pm 0.03$\\
1.52--1.60  & $ 1.67$$\pm$$ 0.02$  & $123$$\pm$$  5$  & $128$$\pm$$  6$  & $ 2.52$$\pm$$ 0.05$  & $ 11.4$$\pm$$  1.3$  & $ 0.08$$\pm$$ 0.02$  & $ 0.31$$\pm$$ 0.14$  & $1.1$$\pm$$0.2$  & $ 70$$\pm$$ 14$  & $5.1$$\pm$$1.2$ &0.99(134)&$ 1.37\pm 0.03$\\
1.60--1.68  & $ 1.68$$\pm$$ 0.02$  & $120$$\pm$$  4$  & $124$$\pm$$  6$  & $ 2.54$$\pm$$ 0.06$  & $  9.9$$\pm$$  1.1$  & $ 0.08$$\pm$$ 0.02$  & $ 0.31$$\pm$$ 0.14$  & $1.1$$\pm$$0.2$  & $ 74$$\pm$$ 15$  & $4.9$$\pm$$1.2$ &1.18(131)&$ 1.34\pm 0.03$\\
1.68--1.76  & $ 1.65$$\pm$$ 0.03$  & $125$$\pm$$  5$  & $130$$\pm$$  7$  & $ 2.50$$\pm$$ 0.07$  & $ 10.3$$\pm$$  1.4$  & $ 0.07$$\pm$$ 0.03$  & $ 0.34$$\pm$$ 0.11$  & $1.3$$\pm$$0.2$  & $ 87$$\pm$$ 14$  & $5.4$$\pm$$1.3$ &1.23(126)&$ 1.31\pm 0.03$\\
1.76--1.84  & $ 1.67$$\pm$$ 0.03$  & $120$$\pm$$  5$  & $124$$\pm$$  6$  & $ 2.52$$\pm$$ 0.07$  & $  9.0$$\pm$$  1.3$  & $ 0.07$$\pm$$ 0.03$  & $ 0.35$$\pm$$ 0.11$  & $1.4$$\pm$$0.2$  & $ 96$$\pm$$ 16$  & $5.7$$\pm$$1.3$ &0.96(130)&$ 1.27\pm 0.03$\\
1.84--1.92  & $ 1.69$$\pm$$ 0.03$  & $114$$\pm$$  4$  & $115$$\pm$$  6$  & $ 2.60$$\pm$$ 0.08$  & $  7.3$$\pm$$  1.1$  & $ 0.03$$\pm$$ 0.03$  & $ 0.36$$\pm$$ 0.10$  & $1.4$$\pm$$0.2$  & $106$$\pm$$ 17$  & $5.7$$\pm$$1.4$ &0.97(120)&$ 1.23\pm 0.04$\\
1.92--2.00  & $ 1.70$$\pm$$ 0.04$  & $109$$\pm$$  6$  & $112$$\pm$$  9$  & $ 2.61$$\pm$$ 0.14$  & $  6.4^{+  2.1}_{ -1.4}$  & $ 0.04$$\pm$$ 0.05$  & $ 0.35$$\pm$$ 0.12$  & $1.5$$\pm$$0.3$  & $113$$\pm$$ 23$  & $6.5$$\pm$$2.1$ &0.88(111)&$ 1.20\pm 0.06$\\
2.00--2.08  & $ 1.69$$\pm$$ 0.03$  & $114$$\pm$$  5$  & $115$$\pm$$  6$  & $ 2.58$$\pm$$ 0.08$  & $  7.3$$\pm$$  1.2$  & $ 0.03$$\pm$$ 0.03$  & $ 0.35$$\pm$$ 0.09$  & $1.6$$\pm$$0.2$  & $118$$\pm$$ 17$  & $6.2$$\pm$$1.4$ &1.01(122)&$ 1.21\pm 0.03$\\
2.08--2.16  & $ 1.70$$\pm$$ 0.02$  & $112$$\pm$$  4$  & $113$$\pm$$  4$  & $ 2.61$$\pm$$ 0.06$  & $  7.3$$\pm$$  0.8$  & $ 0.02$$\pm$$ 0.02$  & $ 0.34$$\pm$$ 0.08$  & $1.6$$\pm$$0.2$  & $118$$\pm$$ 15$  & $5.9$$\pm$$1.2$ &1.38(128)&$ 1.23\pm 0.02$\\
2.16--2.24  & $ 1.71$$\pm$$ 0.02$  & $110$$\pm$$  4$  & $113$$\pm$$  5$  & $ 2.52$$\pm$$ 0.06$  & $  8.7$$\pm$$  1.2$  & $ 0.05$$\pm$$ 0.02$  & $ 0.33$$\pm$$ 0.08$  & $1.7$$\pm$$0.2$  & $118$$\pm$$ 14$  & $7.1$$\pm$$1.2$ &1.15(131)&$ 1.27\pm 0.03$\\
2.24--2.32  & $ 1.74$$\pm$$ 0.03$  & $106$$\pm$$  4$  & $108$$\pm$$  5$  & $ 2.56$$\pm$$ 0.07$  & $  8.7$$\pm$$  1.2$  & $ 0.03$$\pm$$ 0.02$  & $ 0.40$$\pm$$ 0.08$  & $1.9$$\pm$$0.2$  & $129$$\pm$$ 17$  & $6.2$$\pm$$1.2$ &1.25(127)&$ 1.31\pm 0.03$\\
2.32--2.40  & $ 1.80$$\pm$$ 0.03$  & $ 96$$\pm$$  4$  & $ 98$$\pm$$  5$  & $ 2.59$$\pm$$ 0.08$  & $  8.1$$\pm$$  1.4$  & $ 0.04$$\pm$$ 0.02$  & $ 0.34$$\pm$$ 0.08$  & $2.0$$\pm$$0.2$  & $124$$\pm$$ 16$  & $7.1$$\pm$$1.2$ &1.12(125)&$ 1.34\pm 0.04$\\
2.40--2.48  & $ 1.82$$\pm$$ 0.03$  & $ 95$$\pm$$  4$  & $ 95$$\pm$$  5$  & $ 2.62^{+ 0.05}_{-0.08}$  & $  8.3$$\pm$$  1.4$  & $ 0.02$$\pm$$ 0.02$  & $ 0.39$$\pm$$ 0.08$  & $2.1$$\pm$$0.3$  & $129$$\pm$$ 16$  & $6.3$$\pm$$1.3$ &1.33(123)&$ 1.38\pm 0.04$\\
2.48--2.56  & $ 1.85$$\pm$$ 0.03$  & $ 89$$\pm$$  4$  & $ 91$$\pm$$  5$  & $ 2.61$$\pm$$ 0.09$  & $  8.4$$\pm$$  1.6$  & $ 0.04$$\pm$$ 0.03$  & $ 0.35$$\pm$$ 0.08$  & $2.0$$\pm$$0.2$  & $120$$\pm$$ 15$  & $7.1$$\pm$$1.2$ &1.41(130)&$ 1.41\pm 0.04$\\
2.56--2.64  & $ 1.90$$\pm$$ 0.04$  & $ 84$$\pm$$  4$  & $ 85$$\pm$$  5$  & $ 2.69$$\pm$$ 0.10$  & $  7.4$$\pm$$  1.5$  & $ 0.02^{+ 0.03}$  & $ 0.36$$\pm$$ 0.09$  & $2.1$$\pm$$0.3$  & $118$$\pm$$ 16$  & $7.8$$\pm$$1.3$ &1.30(129)&$ 1.44\pm 0.05$\\
2.64--2.72  & $ 1.93$$\pm$$ 0.04$  & $ 81$$\pm$$  4$  & $ 82$$\pm$$  5$  & $ 2.67$$\pm$$ 0.11$  & $  7.7$$\pm$$  1.7$  & $ 0.04$$\pm$$ 0.03$  & $ 0.35$$\pm$$ 0.09$  & $2.1$$\pm$$0.3$  & $112$$\pm$$ 15$  & $7.1$$\pm$$1.2$ &1.42(128)&$ 1.48\pm 0.05$\\
2.72--2.80  & $ 1.98$$\pm$$ 0.03$  & $ 77$$\pm$$  3$  & $ 77$$\pm$$  4$  & $ 2.77^{+ 0.06}_{-0.11}$  & $  6.7$$\pm$$  1.4$  & $ 0.01^{+ 0.03}$  & $ 0.34$$\pm$$ 0.09$  & $2.1$$\pm$$0.3$  & $112$$\pm$$ 16$  & $6.8$$\pm$$1.2$ &1.41(128)&$ 1.52\pm 0.04$\\
2.80--2.88  & $ 2.00$$\pm$$ 0.04$  & $ 75$$\pm$$  3$  & $ 75^{+  5}_{ -3}$  & $ 2.76$$\pm$$ 0.10$  & $  7.3^{+  1.9}_{ -1.0}$  & $ 0.00^{+ 0.03}$  & $ 0.36$$\pm$$ 0.09$  & $2.2$$\pm$$0.3$  & $111$$\pm$$ 16$  & $6.7$$\pm$$1.2$ &1.41(128)&$ 1.55\pm 0.04$\\
2.88--2.96  & $ 2.02$$\pm$$ 0.05$  & $ 73$$\pm$$  4$  & $ 74$$\pm$$  5$  & $ 2.73$$\pm$$ 0.12$  & $  7.8$$\pm$$  1.9$  & $ 0.03^{+0.03}$  & $ 0.36$$\pm$$ 0.09$  & $2.3$$\pm$$0.3$  & $112$$\pm$$ 16$  & $6.9$$\pm$$1.2$ &1.23(130)&$ 1.59\pm 0.05$\\
2.96--3.04  & $ 2.06^{+ 0.03}_{-0.02}$  & $ 70$$\pm$$  3$  & $ 70^{+  1}_{ -2}$  & $ 2.79$$\pm$$ 0.06$  & $  7.3$$\pm$$  1.2$  & $ 0.00^{+ 0.02}$  & $ 0.33^{+ 0.08}_{-0.12}$  & $2.2$$\pm$$0.3$  & $105$$\pm$$ 14$  & $7.3$$\pm$$1.3$ &1.42(122)&$ 1.63\pm 0.03$\\
3.04--3.12  & $ 2.08$$\pm$$ 0.06$  & $ 68$$\pm$$  4$  & $ 69$$\pm$$  5$  & $ 2.74$$\pm$$ 0.12$  & $  7.9$$\pm$$  2.3$  & $ 0.02^{+ 0.03}$  & $ 0.38$$\pm$$ 0.10$  & $2.3$$\pm$$0.3$  & $105$$\pm$$ 19$  & $7.2$$\pm$$1.3$ &1.33(123)&$ 1.66\pm 0.06$\\
3.12--3.20  & $ 2.15$$\pm$$ 0.04$  & $ 62$$\pm$$  3$  & $ 62$$\pm$$  3$  & $ 2.88^{+ 0.08}_{-0.12}$  & $  6.0^{+  1.7}_{ -1.2}$  & $ 0.00^{+ 0.03}$  & $ 0.35$$\pm$$ 0.10$  & $2.2$$\pm$$0.3$  & $ 97$$\pm$$ 16$  & $7.3$$\pm$$1.3$ &1.25(125)&$ 1.68\pm 0.05$\\
3.20--3.40  & $ 2.17$$\pm$$ 0.05$  & $ 62$$\pm$$  3$  & $ 62$$\pm$$  4$  & $ 2.85^{+ 0.06}_{-0.12}$  & $  7.0^{+  2.2}_{ -1.2}$  & $ 0.00^{+ 0.03}$  & $ 0.38$$\pm$$ 0.09$  & $2.3$$\pm$$0.3$  & $100$$\pm$$ 17$  & $7.0$$\pm$$1.2$ &1.39(126)&$ 1.75\pm 0.05$\\
3.40--3.60  & $ 2.26$$\pm$$ 0.06$  & $ 57$$\pm$$  4$  & $ 57$$\pm$$  4$  & $ 2.92^{+ 0.11}_{-0.17}$  & $  5.9^{+  2.8}_{ -1.6}$  & $ 0.01^{+ 0.04}$  & $ 0.38$$\pm$$ 0.11$  & $2.3$$\pm$$0.4$  & $ 94$$\pm$$ 16$  & $6.5$$\pm$$1.2$ &1.25(131)&$ 1.86\pm 0.08$\\
3.60--3.80  & $ 2.36^{+ 0.05}_{-0.09}$  & $ 51$$\pm$$  3$  & $ 51^{+  6}_{ -3}$  & $ 3.03^{+ 0.13}_{-0.24}$  & $  4.6^{+  3.6}_{ -1.5}$  & $ 0.00^{+ 0.05}$  & $ 0.40$$\pm$$ 0.12$  & $2.2$$\pm$$0.4$  & $ 84$$\pm$$ 19$  & $6.3$$\pm$$1.2$ &1.33(118)&$ 1.97\pm 0.10$\\
3.80--4.00  & $ 2.39$$\pm$$ 0.12$  & $ 49^{+  8}_{ -5}$  & $ 50^{+  9}_{ -5}$  & $ 2.91$$\pm$$ 0.27$  & $  6.0^{+  6.3}_{ -3.2}$  & $ 0.04^{+0.04}$  & $ 0.40$$\pm$$ 0.13$  & $2.5$$\pm$$0.5$  & $ 88$$\pm$$ 19$  & $6.6$$\pm$$1.3$ &1.24(128)&$ 2.06\pm 0.16$\\
4.00--4.20  & $ 2.54^{+ 0.08}_{-0.16}$  & $ 42^{+  7}_{ -3}$  & $ 43^{+  8}_{ -3}$  & $ 3.16$$\pm$$ 0.35$  & $  3.5^{+  5.8}_{ -0.5}$  & $ 0.01^{+ 0.07}$  & $ 0.41$$\pm$$ 0.13$  & $2.4$$\pm$$0.4$  & $ 83$$\pm$$ 20$  & $5.3$$\pm$$1.3$ &1.38(121)&$ 2.14\pm 0.15$\\
4.20--4.40  & $ 2.62^{+ 0.11}_{-0.22}$  & $ 38^{+  9}_{ -3}$  & $ 39^{+ 10}_{ -4}$  & $ 3.14^{+ 0.60}_{-0.38}$  & $  3.6^{+  8.2}_{ -2.4}$  & $ 0.02^{+ 0.06}$  & $ 0.44$$\pm$$ 0.14$  & $2.6$$\pm$$0.5$  & $ 83$$\pm$$ 20$  & $5.4$$\pm$$1.3$ &1.12(125)&$ 2.23\pm 0.21$\\
4.40--4.80  & $ 2.89^{+ 0.05}_{-0.31}$  & $ 29^{+  5}_{ -1}$  & $ 29^{+ 10}_{ -1}$  & $ 3.49$$\pm$$ 0.71$  & $  1.1^{+  9.3}_{ -0.2}$  & $ 0.00^{+ 0.05}$  & $ 0.44$$\pm$$ 0.13$  & $2.5$$\pm$$0.5$  & $ 77$$\pm$$ 21$  & $4.2$$\pm$$1.2$ &1.02(127)&$ 2.33\pm 0.24$
\enddata 
\tablecomments{See text and Table~\ref{tbl-fitparmod1} for the meaning of each column.}
\end{deluxetable}

\tabletypesize{\scriptsize}
\setlength{\tabcolsep}{0.03in}
\begin{deluxetable}{cccccccccccc}
\tablecaption{HBO fundamental and Upper kHz QPOs\label{tbl-upqpo}}
\tablewidth{0pt}
\tablehead{& \multicolumn{3}{c}{HBO fundamental} & & \multicolumn{3}{c}{Lower kHz QPO} & & \multicolumn{3}{c}{Upper kHz QPO}\\
 \cline{2-4} \cline{6-8} \cline{10-12}
\colhead{$S_{\rm Z}$} & \colhead{Frequency(Hz)} & \colhead{FWHM(Hz)} &\colhead{rms} & & \colhead{Frequency(Hz)} & \colhead{FWHM(Hz)} &\colhead{rms} & & \colhead{Frequency(Hz)} & \colhead{FWHM(Hz)} &\colhead{rms}}
\startdata
0.07--0.24  & $ 22.0$$\pm$$  0.3$  & $  4.3$$\pm$$  1.3$  & $0.032$$\pm$$0.005$ & &  & & & & $622.8$$\pm$$ 28.3$  & $232.3^{+ 67.7}_{-97.0}$  & $0.064$$\pm$$0.012$ \\
0.24--0.32  & $ 24.2$$\pm$$  0.5$  & $  4.8^{+  2.7}_{ -1.6}$  & $0.029$$\pm$$0.006$ & &  & & & & $626.0$$\pm$$ 21.1$  & $149.5^{+101.0}_{-58.1}$  & $0.055$$\pm$$0.010$ \\
0.32--0.40  & $ 26.9$$\pm$$  0.3$  & $  5.7$$\pm$$  1.2$  & $0.035$$\pm$$0.004$ & &  & & & & $683.1^{+ 14.4}_{-20.6}$  & $112.4^{+ 68.5}_{-38.1}$  & $0.042$$\pm$$0.007$ \\
0.40--0.48  & $ 28.4$$\pm$$  0.3$  & $  3.5^{+  1.3}_{ -0.9}$  & $0.024$$\pm$$0.004$ & &  & & & & $690.5$$\pm$$ 17.3$  & $156.9$$\pm$$ 50.1$  & $0.049$$\pm$$0.007$ \\
0.48--0.56  & $ 29.8$$\pm$$  0.5$  & $  4.4$$\pm$$  1.6$  & $0.028$$\pm$$0.005$ & &  & & & & $693.0$$\pm$$ 32.4$  & $143.9^{+111.2}_{-63.2}$  & $0.042$$\pm$$0.010$ \\
0.56--0.64  & $ 34.2$$\pm$$  0.7$  & $  4.3^{+  5.9}_{ -2.3}$  & $0.020^{+0.013}_{-0.005}$ & &  & & & & $741.9$$\pm$$ 13.4$  & $ 79.3^{+ 64.9}_{-28.9}$  & $0.039$$\pm$$0.008$ \\
0.64--0.72  & $ 36.6$$\pm$$  0.4$  & $  7.4$$\pm$$  1.6$  & $0.030$$\pm$$0.003$ & &  & & & & $773.6$$\pm$$ 23.8$  & $205.5$$\pm$$ 85.0$  & $0.047$$\pm$$0.008$ \\
0.72--0.80  & $ 39.6$$\pm$$  0.5$  & $  8.7^{+  2.3}_{ -1.4}$  & $0.028$$\pm$$0.003$ & &  & & & & $797.0$$\pm$$  8.4$  & $108.9$$\pm$$ 25.0$  & $0.040$$\pm$$0.004$ \\
0.80--0.88  & $ 42.3$$\pm$$  0.5$  & $  9.4$$\pm$$  2.1$  & $0.027^{+0.003}_{-0.005}$ && $518.2^{+  8.1}_{ -5.5}$  & $ 22.1^{+ 21.5}_{-12.0}$  & $0.015$$\pm$$0.004$ & & $827.3$$\pm$$ 12.0$  & $123.1$$\pm$$ 35.0$  & $0.036$$\pm$$0.005$  \\
0.88--0.96  & $ 45.1$$\pm$$  0.7$  & $  9.2$$\pm$$  3.0$  & $0.023$$\pm$$0.004$ && $552.6$$\pm$$ 20.3$  & $ 58.1$$\pm$$ 36.0$  & $0.017$$\pm$$0.004$ & & $853.6$$\pm$$ 12.9$  & $110.9$$\pm$$ 35.8$  & $0.032$$\pm$$0.005$  \\
0.96--1.00  & $ 49.6^{+  1.0}_{ -2.9}$  & $  4.5^{+ 12.2}_{ -3.2}$  & $0.014^{+0.013}_{-0.006}$ & &  & & & & $899.3$$\pm$$ 26.2$  & $141.4^{+104.4}_{-65.9}$  & $0.036$$\pm$$0.010$ \\
1.00--1.04  & $ 52.3$$\pm$$  0.9$  & $  4.5^{+  4.3}_{ -2.3}$  & $0.017$$\pm$$0.005$ & & $619.5$$\pm$$ 22.9$  & $ 67.1$$\pm$$ 35.9$  & $0.026$$\pm$$0.007$ & & $891.3^{+ 59.7}_{-17.6}$  & $ 34.6^{+188.2}_{-34.6}$  & $0.018^{+0.016}_{-0.010}$ \\
1.04--1.12  & $ 53.3$$\pm$$  1.7$  & $  2.1^{+ 12.8}_{ -1.9}$  & $0.009^{+0.007}_{-0.004}$ && $653.6$$\pm$$ 12.0$  & $ 48.3^{+ 38.8}_{-21.9}$  & $0.021$$\pm$$0.005$ & & $962.7$$\pm$$ 45.5$  & $174.5$$\pm$$110.5$  & $0.028$$\pm$$0.013$ \\
1.12--1.20  & $ 57.2$$\pm$$  0.9$  & $  5.9$$\pm$$  2.7$  & $0.015$$\pm$$0.003$ && $690.3$$\pm$$ 16.1$  & $ 70.9$$\pm$$ 28.6$  & $0.024$$\pm$$0.004$ & & $947.1$$\pm$$ 16.0$  & $ 65.0^{+ 46.7}_{-31.6}$  & $0.021$$\pm$$0.006$ \\
1.20--1.28  & $ 59.6$$\pm$$  0.7$  & $  3.3$$\pm$$  2.4$  & $0.014$$\pm$$0.003$ &\\
1.28--1.36  & $ 60.0$$\pm$$  0.7$  & $  4.0$$\pm$$  3.2$  & $0.018^{+0.004}_{-0.007}$ &\\
1.36--1.44  & $ 60.9$$\pm$$  1.8$  & $  0.9^{+ 19.5}_{ -0.8}$  & $0.011$$\pm$$0.004$ &\\
1.44--1.52  & $ 56.4$$\pm$$  5.7$  & $ 15.0$  & $0.017$$\pm$$0.004$ &\\
1.52--1.60  & $ 53.3$$\pm$$  4.0$  & $ 16.4^{+ 11.7}_{ -6.7}$  & $0.023$$\pm$$0.005$ &\\
1.60--1.68  & $ 55.3^{+  4.9}_{ -9.1}$  & $ 15.0$  & $0.018$$\pm$$0.005$ &\\
1.68--1.76  & $ 54.5^{+  2.7}_{ -4.4}$  & $ 13.7^{+ 16.7}_{ -8.1}$  & $0.027$$\pm$$0.008$ &\\
1.76--1.84  & $ 53.8^{+ 10.5}_{ -7.2}$  & $ 15.0$  & $0.022$$\pm$$0.006$ &\\
1.84--1.92  & $ 51.0^{+  4.6}_{ -8.1}$  & $ 15.0$  & $0.020$$\pm$$0.006$ &\\
1.92--2.00  & $ 44.5^{+  5.0}_{ -7.6}$  & $ 15.0$  & $0.032$$\pm$$0.007$ &
\enddata 
\end{deluxetable}

\end{document}